\documentclass[smallextended]{svjour3}       
\smartqed  
\usepackage{graphicx}
\usepackage[figuresright]{rotating} 
\usepackage{caption}
\usepackage{subcaption}
\usepackage{floatrow}
\usepackage[dvipsnames]{xcolor}
\usepackage{graphicx,amssymb,amsmath}
\usepackage{amsmath,stmaryrd}
\usepackage{natbib}
\graphicspath{{./}{figs/}}
\newcommand{\figref}[1]{Fig.~\ref{fig:#1}}
\newcommand{\secref}[1]{Section~\ref{sec:#1}}
\newcommand{\subsecref}[1]{Section~\ref{subsec:#1}}

\newcommand{\propref}[1]{Proposition~\ref{prop:#1}}

\newcommand{\R}{\mbox{$\mathbb{R}$}}

\newcommand{\E}{\mbox{$\mathbb{E}$}}
\newcommand{\N}{\mbox{$\mathbb{N}$}}
\newcommand{\cD}{\mbox{$\cal D$}}
\newcommand{\cF}{\mbox{$\cal F$}}
\newcommand{\PD}{\mbox{$\sf D$}}
\let\hat\widehat
\let\tilde\widetilde
\usepackage{color}

\journalname{Paper in Submission}
\begin{document}

\bibliographystyle{agsm}

\title{Functional Summaries of Persistence Diagrams
}


\author{Eric Berry \and Yen-Chi Chen \and Jessi Cisewski-Kehe \and Brittany Terese Fasy
}


\institute{E. Berry \at Department of Mathematical Sciences, Montana State University
              \email{eric.berry1@montana.edu}
           \and
           Y. Chen \at
              Department of Statistics, University of Washington
              \email{yenchic@uw.edu}
                       \and
           J. Cisewski-Kehe \at
              Department of Statistics and Data Science, Yale University
              \email{jessica.cisewski@yale.edu}
                       \and
           B. T. Fasy \at
          School of Computing and Department of Mathematical Sciences, Montana State University
              \email{brittany@fasy.us}
}

\date{Received: date / Accepted: date}

\maketitle

\begin{abstract}
One of the primary areas of interest in applied algebraic topology is persistent homology, and, more specifically, the persistence
diagram. Persistence diagrams have also become objects of interest in topological data analysis.  However, persistence diagrams do not naturally lend themselves to
statistical goals, such as inferring certain population characteristics,
because their complicated structure makes common algebraic operations--such as addition, division, and
multiplication-- challenging (e.g., the mean might not be unique).  To bypass these issues, several
functional summaries of persistence diagrams have been
proposed in the literature (e.g. landscape and silhouette functions).
The problem of analyzing a set of
persistence diagrams then becomes the problem of analyzing a set of functions, which is a topic
that has been studied for decades in statistics.
First, we review the various functional summaries in the literature and propose a unified framework for the functional summaries. Then, we generalize the definition of persistence landscape functions, establish several theoretical properties of the persistence functional summaries, and demonstrate and discuss their performance in the context of classification using simulated prostate cancer histology data, and two-sample hypothesis tests comparing human and monkey fibrin images, after developing a simulation study using a new data generator we call \emph{the Pickup Sticks} Simulator (STIX).
\end{abstract}

\section{Introduction} \label{sec:intro}

Topological data analysis (TDA) seeks to understand and characterize topological
features of data.  In particular, persistent homology provides a framework for
analyzing the topological connectivity of a dataset at different scales.
Persistent homology has drawn interest in
applied mathematics \citep{Edelsbrunner:2002qf, CarlssonEtAl2004, Zomorodian2005,
Ghrist2008, Carlsson2009}, computer science and machine learning
\citep{adams2015persistence}, statistics \citep{Worsley1996, AdlerEtAl2010,
chazal2014stochastic, fasy2014confidence, Turner:2014ty, bubenik2015statistical,
chen2015statistical}, and the applied sciences
\citep{SousbieEtAl2011, Van-de-Weygaert:2011aa, cisewski2014non, Singh:2014aa,
bendich2016persistent}.
The interest in persistent homology is due, at least in part, to its ability to
extract summaries of data that are otherwise missed by traditional data analysis methods.
Persistent homology gives a multi-scaled way to view data.  Different features
(in particular, the homology generators) are tracked across the scales, resulting in
an object known as the \emph{persistence diagram}, which is a multi-set of
birth-death pairs (indicating the birth and death of the homology generators).
While
persistence diagrams contain potentially useful information about a dataset,
they are not easy objects to use directly in machine learning and statistical
settings, so work has been carried out to
transform persistence diagrams into one- or two-dimensional functional summaries
and vectorized representations.
In this article,
we review functional summaries of persistence diagrams,
develop
a unified framework for the functional summaries along with proposing a generalization of a popular functional summary (persistence landscape functions), discuss various ways the functional summaries can
be used, and compare the results on simulated and real datasets.
Below, we introduce two examples that will be considered in subsequent sections.

\paragraph{Example: Prostate Cancer Histology.}
\begin{figure}
 \centering
    \begin{subfigure}{0.45\textwidth}
        \centering
        \includegraphics[height=1.5in]{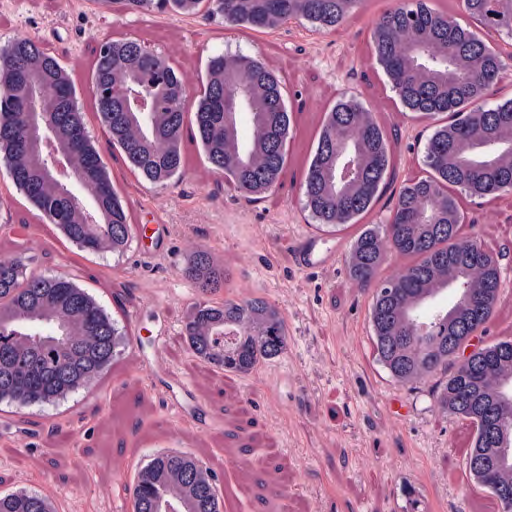}
        \caption{Gleason Grade 3}\label{subfig:grade3}
    \end{subfigure}
    \begin{subfigure}{0.45\textwidth}
        \centering
        \includegraphics[height=1.5in]{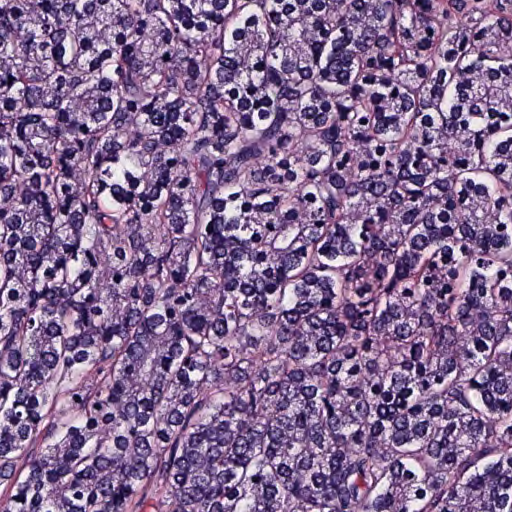}
        \caption{Gleason Grade 5}\label{subfig:grade5}
    \end{subfigure}
    \caption{Sample ROIs from regions graded Gleason Grade $3$ (left) and
    Gleason Grade $5$ (right).  The dark purple are nuclei, and can easily be
    segmented using software such as
    CellProfiler\citep{lamprecht2007cellprofiler}.  In the
    lower-Gleason grades, the nuclei form round circles surrounding glands.  As
    the cancer progresses, these circles become irregular in shape and
    eventually break open, losing all structure.}\label{fig:realslides}
\end{figure}
Prostate cancer (PCa) is the second most common cancer in men worldwide with an
estimated 1.1 million cases diagnosed in 2012~\citep{center_international_2012,
ferlay_cancer_2015}.
Prostate cancer diagnosis involves histological
classification of hematoxylin and eosin (H\&E) prepared slides of a prostate
biopsy, such as the region of interest (ROI) from a slide scan shown in \figref{realslides}.
Slides are classified into five grades based on glandular architectural features
in the Gleason Grading System;
a primary and a secondary grade are assigned, with higher grades
corresponding to increasingly poor prognostic
outcomes~\citep{humphrey_gleason_2004}.  Initially developed in the 1960s, the  Gleason grading system  and its recently introduced refinement, the
Grade Groups~\citep{epstein_2014_2016},
are the most powerful predictors
of prognostic outcome in prostate cancer.  However, the system suffers from high intra- and inter-observer
variability due to the subjective nature of the grading scheme~\citep{engers_reproducibility_2007, truesdale_gleason_2011,
abdollahi_inter/intra-observer_2012, goodman_frequency_2012, helpap_improving_2012,
truong_development_2013, evans_gleason_2016}.
The Gleason grading
system and Grade Groups both rely exclusively on architectural patterns of
carcinoma cells for histological classification; these architectural patterns
provide an opportunity to use topological data analysis.
Through a simulation study (\subsecref{gleasonExper}), we present a technique for classification of
histology slides based on functional summaries of persistence diagrams so that prevalent patterns can be revealed in order to assist a
pathologist or oncologist in finding glandular architectural patterns that can
be used to assign a Gleason Score or Grade~Group.

\paragraph{Example: Fibrin Network Data.}

\begin{figure}
    \centering
        \includegraphics[height=2.5in]{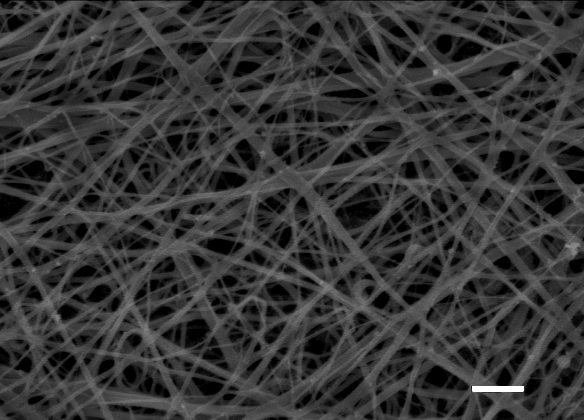}
    \caption{Sample of human fibrin network from Figure 1a of \cite{Pretorius:2009aa}, where the white scale bar is 1 $\mu$m.} \label{fig:fibrin_human_full}
\end{figure}

Complicated spatial structures are common in biological data (e.g., fibrin clots, fibroblasts), but are difficult to quantitatively analyze without losing important information.
In particular, the coagulation cascade culminates in the web-like structure of a
fibrin network.  \figref{fibrin_human_full} displays a sample of a human fibrin network collected using a scanning electron microscope; the image are from \cite{Pretorius:2009aa}.
Features of these structures have numerous implications for vascular diseases like hemophilia and thrombosis \citep{Campbell:2009aa, Pretorius:2009aa}.

One of the primary goals of \cite{Pretorius:2009aa} was to compare platelet and
fibrin networks of humans and eight other animal species, and carry out an
inference procedure to determine if the differences are statistically
significant.  They focused their comparisons on the thickness of the fibrin
fibers and grouped them into thin, intermediate, and thick fibers, requiring
measurements of individual fibers within the collection of fibrin networks.  To
measure the fibers, they randomly selected 100 fibers within the sampled and
imaged fibrin networks, and then measured the diameters of the fibers.  Next,
they assigned the 100 fibers from each fibrin network into one of the
three clusters based on the measured diameter, and then did comparisons between
the fiber classes of the different species.  While they focused on fiber
\emph{thickness}, we compare the \emph{topological structure} of the fibrin networks using functional summaries of persistence diagrams.   In \secref{stix_sim}, we develop a data-generating model that mimics some of the characteristics of these spatially complex data, including allowing for varying thicknesses of the strands, which we call \emph{the Pickup Sticks Simulator} (STIX).  Before carrying out the hypothesis tests on the real fibrin data, we run a simulation study on the STIX data where we know the ground truth and can verify the methodology.

\paragraph{Additional Applications.}
Though we focus on classification and hypothesis tests for prostate cancer histology slides and fibrin networks,
the methods we propose can have analogous use in other areas of science.
For example, the
complicated spatial structure of fibrin networks is similar to the distribution
of matter in the Universe, often referred to as the \emph{Cosmic Web}~\citep{Worsley1996, SousbieEtAl2011, Van-de-Weygaert:2011aa, htesting}, and to brain artery trees \citep{bendich2016persistent,biscio2016accumulated}.

\paragraph{Main Contributions.}

First, a review is provided of the
functional summaries
from the literature.  Then, we propose a unified framework for the functional
summaries and generalize the definition of one popular type of persistence
functional summary, the persistence landscape function, to a broader class of
functions that allows for different kernels and bandwidths providing additional
flexibility for the practitioner.  Theoretical properties of the persistence
functional summaries are established, putting these functional summaries on a
solid foundation for various types of statistical inference and methodology.
Then, several of the popular functional summaries, along with the proposed
generalized landscape function, are used in two different applications in order
to compare their performance.  The first is using simulated prostate cancer
histology data, where the data are classified according to simulated
architectural morphology.
The second application carries out two-sample hypothesis tests comparing
human and monkey fibrin images, which is particularly challenging because the
data have a complicated, web-like, spatial structure.  In order to evaluate the
performance in a more controlled setting, we develop a new data-generating
mechanism, STIX.  STIX
generates images that have similar features to the fibrin data, but are also
interesting objects of study on their~own.

\section{Functional Summaries of Persistence Diagrams}

Before introducing functional summaries of persistence diagrams, we provide an
introduction to persistence diagrams themselves, along with some examples of the
types of inference one may want to do with a set, or sets, of persistence~diagrams.

\subsection{Persistence Diagrams}

In TDA,
persistence diagrams \citep{cohen2007stability,edelsbrunner2012persistent,wasserman2016topological}
provide a useful way
to summarize the topological structure of a point-cloud of data or a function\footnote{Actually, as long as we have a filtration, we
can define a persistence diagram. }.
In this introduction, we focus on the function-based filtrations for persistence diagrams, but other types of filtrations can be used.  For more details on filtrations over point clouds, we suggest \cite{Zhu:2013hl} or \cite{Ghrist:2014rz}.
Given a function $f:\mathcal{X}\to \R$,
where $\mathcal{X}$ is a topological space,
we define the upper\footnote{Some literature considers the lower level set $f^{-1}((-\infty, \lambda))$.
Both definitions are valid and here we use upper level sets because they yield a very straight forward
definition when we consider the image data or certain functions estimated from a point cloud, such as kernel density estimates \citep{wasserman2006all}.
} $\lambda$-level set of $f$ as
$$
L_\lambda = \{x: f(x)\geq \lambda\} =f^{-1}([\lambda,\infty)).
$$
For any two level thresholds $\lambda_1>\lambda_2$,
the corresponding level sets satisfy $L_{\lambda_1}\subset L_{\lambda_2}$.
Thus,
the collection of level sets $\mathcal{L} = \bigcup_{\lambda}\{L_{\lambda}\}$
forms a filtration with the level as the index set.

For each level set $L_{\lambda}$, its topological features are captured through the generators of its homology
groups.
Informally, the $0$-th order homology groups ($0$-th order topological feature)
capture the connected components, the $1$-st order homology groups capture regions forming a loop structure,
and $2$-nd order homology groups capture regions forming a void structure.
For the formal definition of homology groups, we refer readers to \cite{EdelsbrunnerHarer2008} and \cite{edelsbrunner2012persistent}.
When we decrease the level $\lambda$,
new generators for the homology groups may be created (e.g., the formation of new components),
existing generators may merge together (e.g., two connected components joining together), and existing
generators may be eliminated (e.g., a loop getting filled in).
The level at which a generator is created is called its birth time
and the level at which a generator is eliminated,
or merges with another generator that has an earlier birth time,
is called its death time.
Thus, for every generator in $\mathcal{L}$,
there are three characteristics: homology order, birth time, and death time.

The persistence diagram is the collection of all these triplets of the filtration formed by the given function $f$.
Thus, if a function's filtration has $|D|$ generators, then
its persistence diagram is
$$
\PD = \{(r_j,b_j,d_j): j =1,\cdots,|D|\},
$$
where $r_j,b_j,$ and $d_j$ are the homology order, birth time, and death time of the $j$-th generator, respectively, and the norm $|D|$ denotes the number
of off-diagonal elements in the persistence diagram $\PD$.

\subsubsection{Example: constructing a persistence diagram from a dataset}
There are many ways of constructing a persistence diagram from a dataset.
If the data are images, functions, or fields evaluated on a grid,
then the construction of the corresponding persistence diagrams is straight forward --
we just consider the pixels, or grid points,
whose value is above a given level and vary such level to construct a filtration.

\begin{figure}
    \begin{subfigure}{0.225\textwidth}
        \centering
        \includegraphics[height=1.1in]{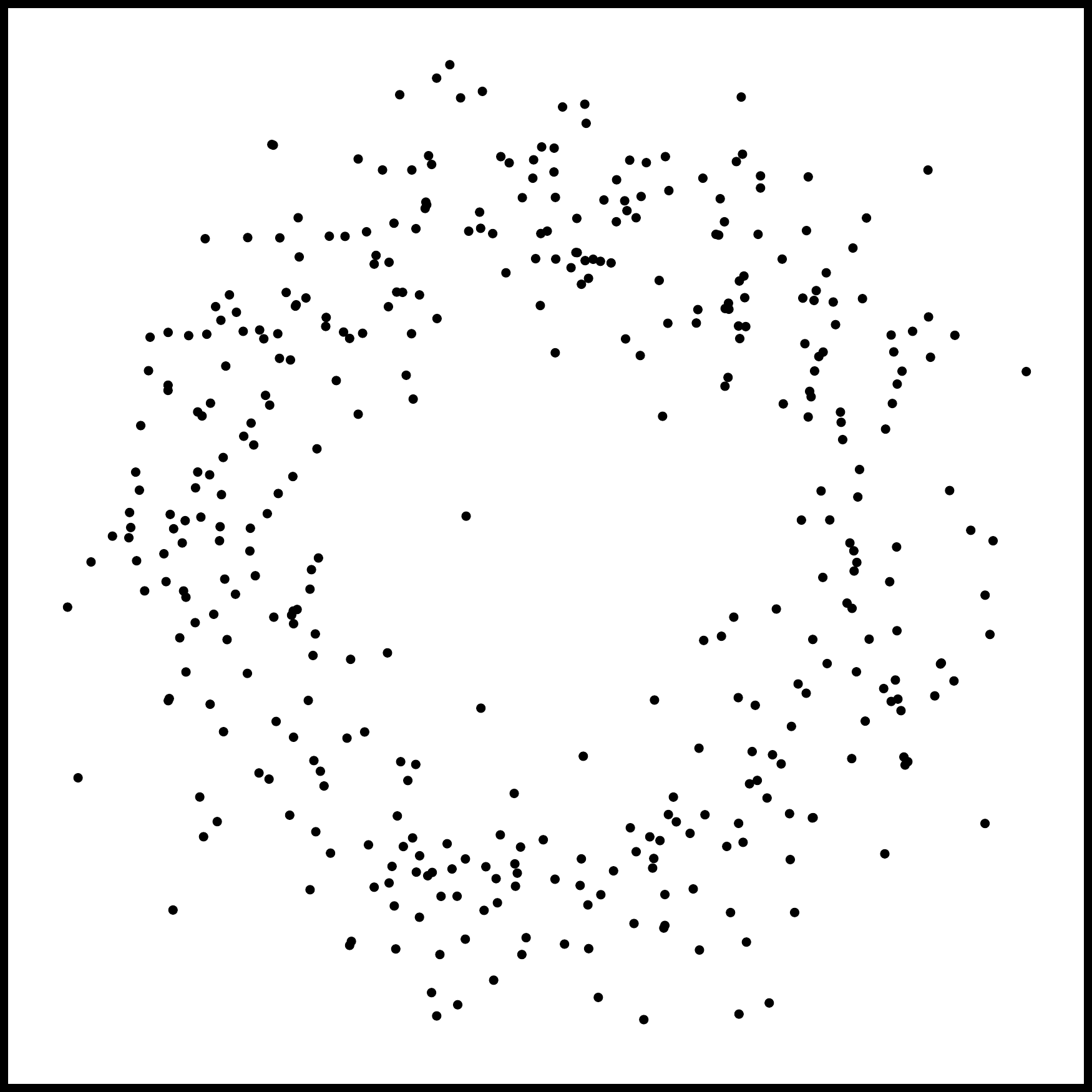}
        \caption{Dataset}\label{subfig:ex01_1}
    \end{subfigure}
    \begin{subfigure}{0.225\textwidth}
        \centering
        \includegraphics[height=1.1in]{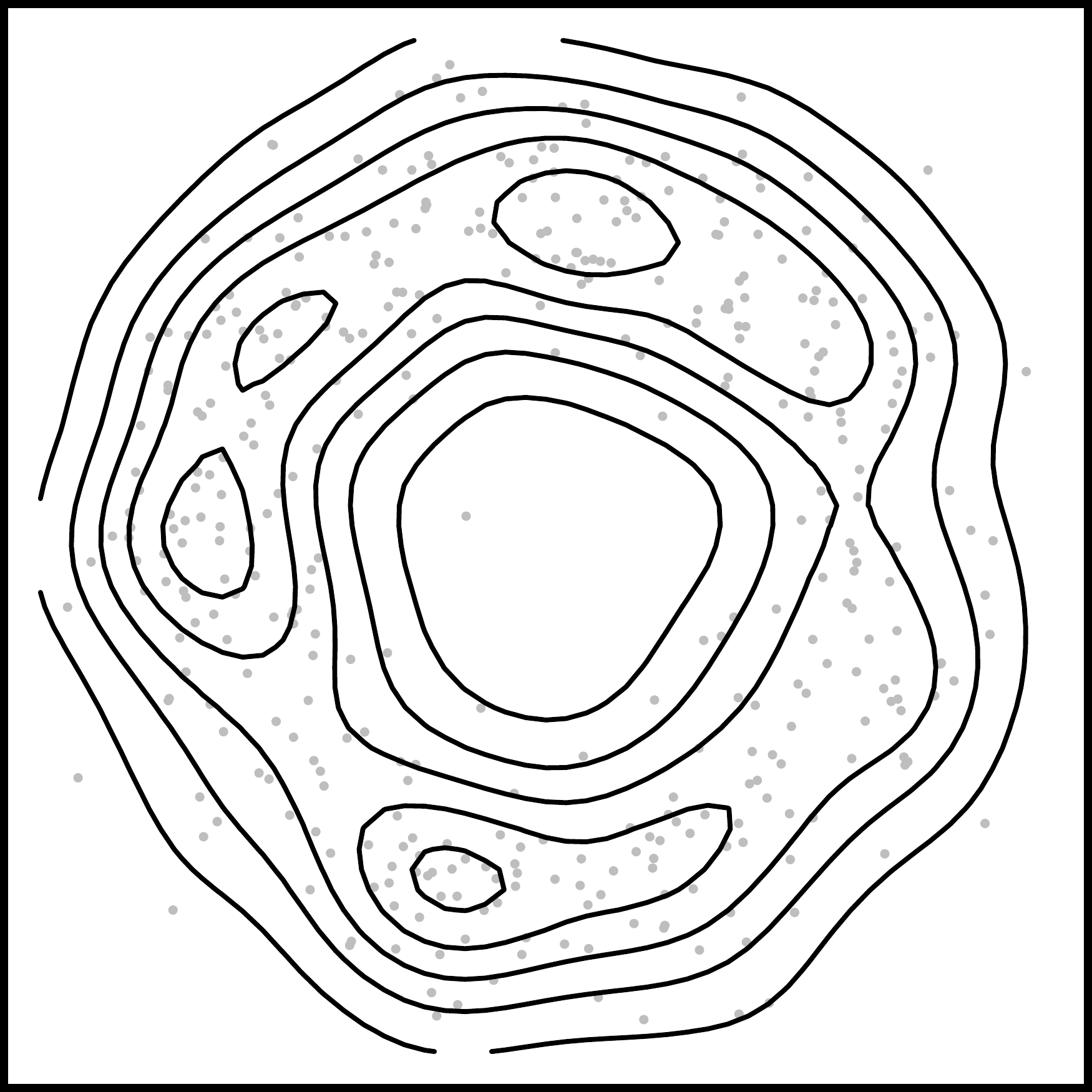}
        \caption{KDE}\label{subfig:ex01_1}
    \end{subfigure}
    \begin{subfigure}{0.225\textwidth}
        \centering
        \includegraphics[height=1.1in]{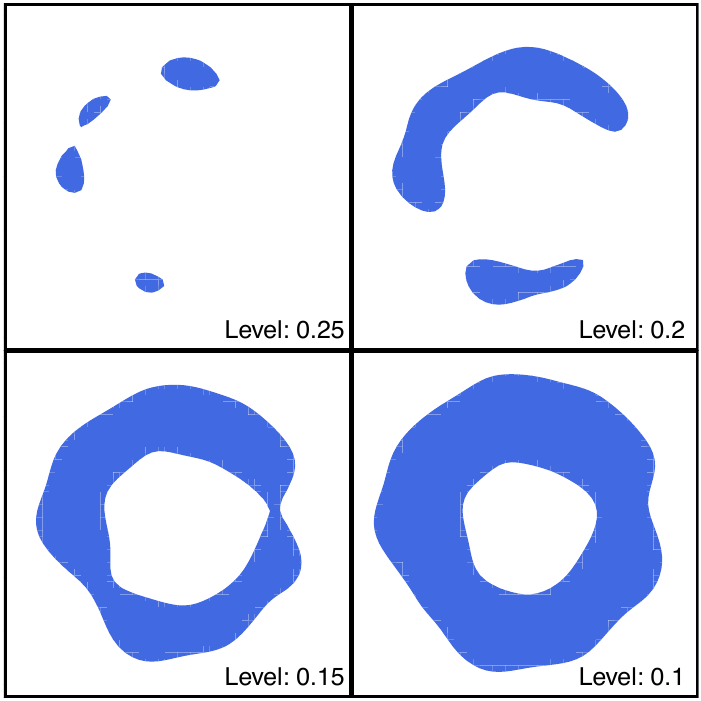}
        \caption{Level Set}\label{subfig:ex01_1}
    \end{subfigure}
    \begin{subfigure}{0.225\textwidth}
        \centering
        \includegraphics[height=1.1in]{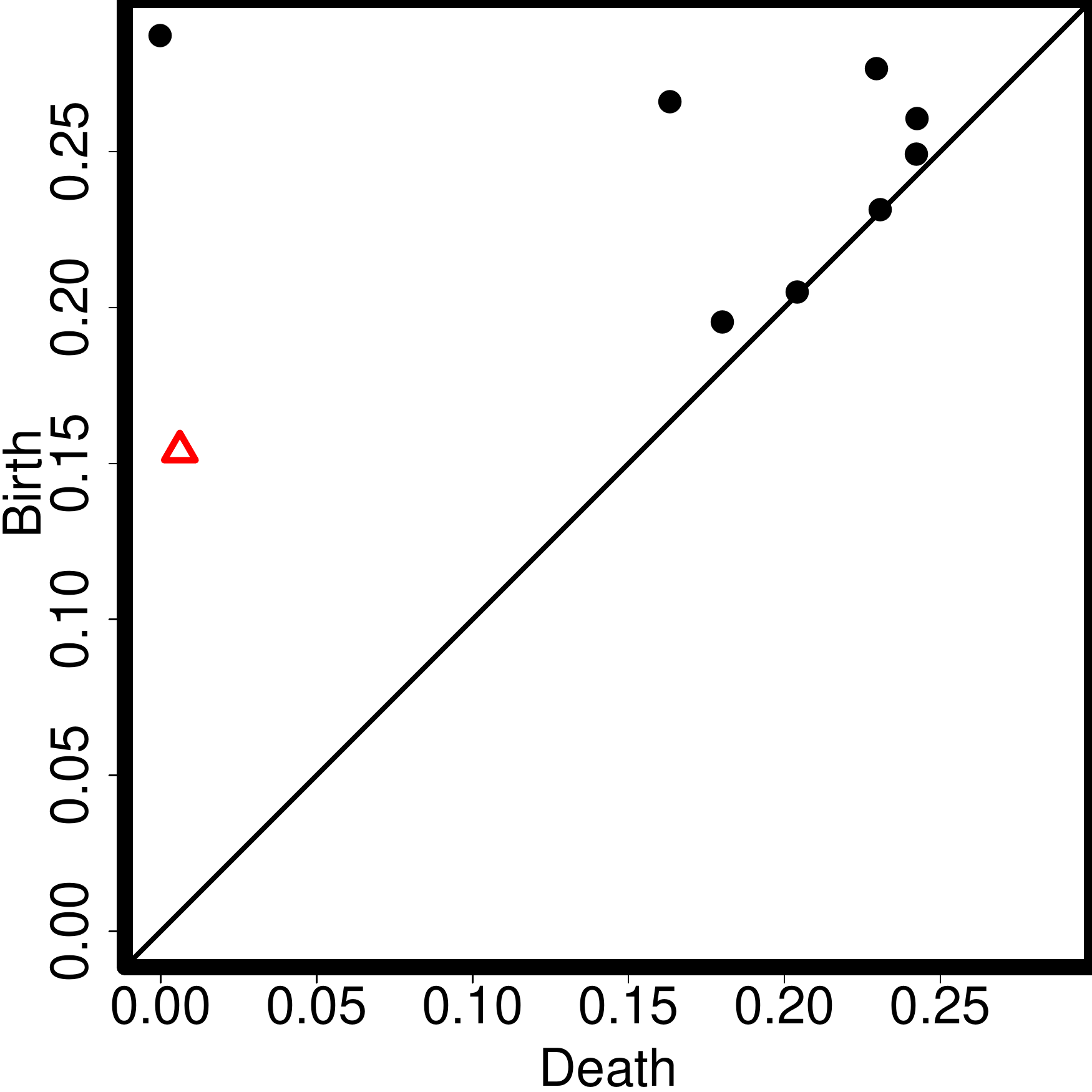}
        \caption{PD}\label{subfig:ex01_1}
    \end{subfigure}

\caption{
The construction of a persistence diagram from the single dataset (points) displayed in (a).
We construct the persistence diagram (PD, d) using level sets (e.g., c) of its kernel density estimator (b).
}
\label{fig:PDex01}
\end{figure}

When the data are a collection of points, the construction of a persistence diagram
depends on how the function that generates the underlying filtration is constructed.
Here we illustrate the construction using an estimator of the underlying density function.
We use a kernel density estimate (KDE) to estimate the underlying probability density function
that generates this data
and construct the filtration using the (upper) level set of the KDE \citep{fasy2014confidence,wasserman2016topological}.
This procedure is summarized in \figref{PDex01},
where we obtain a persistence diagram for the estimated density function
of the given 2D point clouds.
Formally, let $X_1,\cdots,X_n\in \R^d$ be the observed values for a single dataset.
The KDE is
$\hat{p}_h(x) = \frac{1}{nh^d}\sum_{i=1}^n K\left(\frac{\|X_i-x\|}{h}\right),$
where $K(x)$ is a smooth function called the kernel function (e.g. a Gaussian kernel) and $h>0$ is a quantity called the smoothing
bandwidth that controls the amount of smoothing \citep{wasserman2006all}.
Using $\hat{p}_h$ and its level sets, we
then construct the persistence diagram which contains topological information about the underlying density function.

\subsection{Modeling multiple persistence diagrams}

In this paper, we focus on the case where multiple persistence diagrams,
$
\PD_1,\dots, \PD_n,
$
are observed
and we want to perform statistical analysis (e.g., estimating population quantities, classification, hypothesis testing) over these diagrams.
In addition to the prostate cancer histology data and fibrin network data
mentioned previously,
there are many other situations
where this framework is useful, such as when
analyzing brain artery trees \citep{bendich2016persistent,biscio2016accumulated}, investigating the large-scale structure of the Universe using cosmological simulations \citep{htesting},
or studying shape data \citep{Turner:2014ty,crawford2016topological}.
A common assumption
in these examples
is that the persistence diagrams, $\PD_1,\dots, \PD_n $,
are generated from the same population. Thus,
we can model this procedure as the case where there exists a distribution of
persistence diagrams $\mathcal{P}$
\citep{mileyko2011probability} such that
$\PD_1,\dots, \PD_n $ are independent and identically distributed from~$\mathcal{P}$.

Unfortunately, persistence diagrams are not easy objects to work with.
Even if a sample of persistence diagrams are from the same distribution,
each of them may have different numbers of topological features, and those features have a complicated covariance.
This makes it difficult to carry out common algebraic operations such as
addition, division, and multiplication, hence computing statistical summaries
such as means and medians is challenging \citep{Turner:2013fk, Turner:2014kq}.
In the multiple diagram setting,
a simple but elegant approach is to summarize each diagram by a function
and then analyze the diagrams by comparing their corresponding functions.
Because functions are well-defined objects, and statistical analysis
over functions are well-studied,
analyzing these functional summaries is a much easier task than directly studying the diagrams.

The proposed functional summaries use
either a univariate or a bivariate function
function
to summarize the persistence diagram, but, in general, functions with more variables can be used.
A brief review of the existing functional summaries of persistence diagrams is provided next, before giving more details on the proposed functional summaries.
Let $\cF$ be a collection of functions.
Then, a functional summary of a persistence diagram, $\mathbb{F}(\PD)$, is a map
between $\cD$ and $\cF$. i.e.,
$$
\mathbb{F}:\cD \to \cF.
$$

Using a functional summary $\mathbb{F}$,
the random diagrams $\PD_1,\cdots,\PD_n$ become random functions
$F_1 = \mathbb{F}(\PD_1),\cdots,F_n = \mathbb{F}(\PD_n)$.
Moreover, since these random diagrams are from the same distribution, the corresponding functional summaries also come from the same distribution (of functions), $\mathcal{P}_F$:
\begin{equation}
F_1 ,\cdots,F_n \overset{i.i.d.}{\sim} \mathcal{P}_F. \label{eq:model}
\end{equation}
The above characterizes the statistical model for using functional summaries to analyze persistence diagrams.
Except where noted, we focus on topological features of the same homological dimension.

\subsection{Review: Functional Summaries}
We review several
functional summaries that have been proposed in
existing~literature.

\subsubsection{Persistence Landscape}
\begin{figure}
\centering
\includegraphics[width=\textwidth]{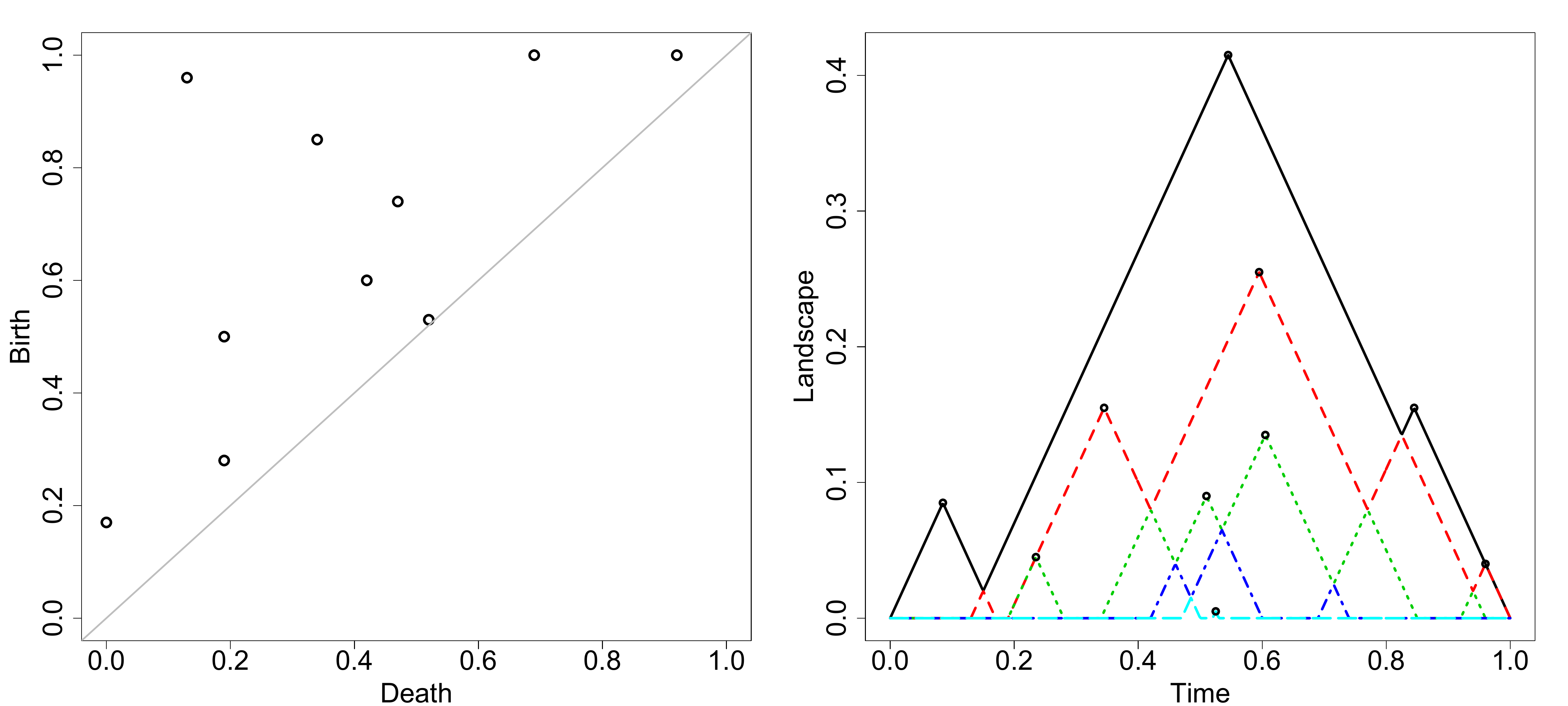}
\caption{Illustration of a persistence landscape function.  The left plot is a persistence diagram and the right plot is its corresponding persistence landscape function; each color and line style represents a different $\mathbb{F}_{k}$ with $\mathbb{F}_{1}$ as the solid black line, $\mathbb{F}_{2}$ as the dashed red line, etc.
} \label{fig:landscape}
\end{figure}

The persistence landscape function is a popular univariate functional summary of a persistence diagram \citep{chazal2014stochastic,bubenik2015statistical}.  To produce a persistence landscape function, or simply referred to as a landscape function, the persistence diagram is rotated clockwise 45 degrees, and then isosceles right triangles are drawn from each feature of a particular homology dimension (where the right angle vertex is the homology feature).  From the collection of isosceles right triangles, individual functions are traced out where the first landscape function is the point-wise maximum of all the triangles drawn.  More specifically, a persistence landscape is a collection of univariate functional summaries $\mathbb{F}_k:\cD \to \cF$
such that for each $k\in\N$,
\begin{equation}
\mathbb{F}_{k}(\PD;t) = \underset{j=1,\cdots, |\PD|}{\sf kmax} \{\Lambda_j(t)\}, \label{eq:landscape}
\end{equation}
where $\Lambda_j(t) = \min(t-b_j,d_j-t)_+$, ${\sf kmax}$ is a function selecting the $k$-th largest value, and the norm $|\PD|$ denotes the number
of off-diagonal elements in the persistence diagram $\PD$.  An illustration of
this is displayed in \figref{landscape}.

\begin{center}
\end{center}

\subsubsection{Persistence Silhouette}
Persistence silhouette functions were introduced in
 \cite{chazal2014stochastic} as a modification of the persistence landscape function.
The persistence silhouette function maps the persistence diagram to a function $\cD \to \R$, as opposed to
$\cD  \to \R \times \N$ with landscape functions,
by combining all the layers of the landscapes functions into a single, average function.
The persistence silhouette is a function $\mathbb{F}:\cD\to \cF_1$ such that
$$
\mathbb{F}(\PD;t) = \frac{\sum_{j = 1}^{\left|\PD\right|} \omega(d_j-b_j) \Lambda_j(t)}{\sum_{k=1}^{\left|\PD\right|}\omega(d_k-b_k)},
$$
where $\omega(d_j-b_j)$ is a weight function based on the persistence of the features.  For example, the weight function $\omega(d_j-b_j) = | d_j - b_j|^p$ could be used, where $p$ is a tuning parameter that has to be selected.  Larger values of $p$ put more weight on the features with longer lifetimes, and smaller values of $p$ emphasize the features with shorter lifetimes.

\subsubsection{Accumulative Persistence Function}
The accumulative persistence function (APF), introduced in
 \cite{biscio2016accumulated},
is a univariate functional summaries
$\mathbb{F}:\cD\to \cF_1$ such that
$$
\mathbb{F}(\PD;t) = \sum_{j=1}^{\left|\PD\right|}(d_j-b_j) \cdot I(d_j+b_j\leq 2t).
$$
The APF behaves like a cumulative distribution function in that it is a non-decreasing function.

\subsubsection{Persistence Intensity and Image}
The persistence intensity is a bivariate functional summary \citep{chen2015statistical}, where $\mathbb{F}: \cD\to \cF_2$ is a map to a bivariate function defined as
\begin{align}
\mathbb{F}(\PD;t,s) =
&\frac{1}{\left|\PD\right|}\sum_{j=1}^{\left|\PD\right|}\omega(d_j-b_j) \cdot K\left(\frac{\sqrt{|b_j-t|^2+|d_j-s|^2}}{h}\right), \label{eq:intensity}
\end{align}
where $\omega$ is a weight function and $K$ is a kernel function with smoothing
parameter $h$.
The persistence intensity replaces points on the persistence diagrams
by a smooth localized function that is determined by the kernel function $L$ (the amount of smoothing is determined by
the parameter $h$)
and apply a weighted sum over all points to form an intensity function
such that, for example,
points far away from the diagonal (features that are more persistent)
are given higher weights.
This leads to a bivariate function that summarizes a persistence diagram.

Note that
if we evaluate the bivariate functional summary on a grid and view it as an image,
we obtain the persistence image introduced in \cite{adams2015persistence}.
The persistence image framework of \cite{adams2015persistence} includes vectorizing the image making it suitable for
many machine learning and statistical methods.\footnote{\cite{adams2015persistence} also presented conditions under which persistence images are stable with respect to changes in the corresponding persistence diagrams.}

\begin{remark}
When we can have an extra scale parameter for each persistence diagram,
we can construct a special bivariate functional summary called the
\emph{persistence flamelet}.
The persistence flamelet \citep{padellini2017persistence} is a bivariate functional summary that combines
the persistence landscape and the scale parameter.
For instance, the persistence diagram may be constructed from
a KDE of a point could making the smoothing bandwidth $h$ in the KDE the scale parameter.
In this case, the persistence landscape also depends on the scale parameter $h$.
The persistence flamelet is a bivariate function of $(t,h)$ such that for each fixed $h$, the persistence flamelet
is the persistence landscape.
\end{remark}

\begin{remark}
In addition to functional summaries of persistence diagrams, one might want
\emph{vector-valued summaries} for input
to machine learning algorithms, for instance. In \cite{adcock2016ring}, the authors define an algebra of functions
on the space of persistence bar codes. Furthermore, they identify the generators of this algebra,
and show the utility of these
functions through applications of classification of data using machine learning.
In addition, using the maximum persistence alone has proven successful in
certain settings \citep{khasawneh2014exploring,perea2015sliding}.
\end{remark}

\section{Generalized Landscape Functions}

As noted previously, landscape functions result from a persistence diagram that
is rotated, and then each feature $j$ on the diagram produces an isosceles right
triangle, $\Lambda_j$.  However, one need not be restricted to these special
triangles when computing one-dimensional functional summaries of persistence
diagrams.  We propose an expanded class of persistence landscape functions,
called \emph{generalized persistence landscape functions}.  With these
generalized landscapes, one can substitute different kernels (e.g. tricube,
Epanechnikov) and different bandwidths in order to better adapt to the details
in the persistence diagram.  (See \cite{wasserman2006all} for a general
discussion about kernel functions and bandwidth selection.)  The triangle kernel
used in the original landscape function definition can also be used, except the
width of the base of the triangle can be adjusted, which could be analogously thought of as adjusting the angles of the triangle.

More specifically, the change is in the form of $\Lambda_j$ from Equation~\eqref{eq:landscape}, where the generalized landscapes use
\[
\widetilde{\Lambda}_j(t; h) = \begin{cases}
         \frac{y_j}{K_h(0)} K_h\left({t - x_j}\right),  & \text{for } | \frac{t - x_j}{h} | \leq 1 \\
        0, & \text{otherwise, }
        \end{cases}
\]
where $K_h$ is the kernel function with $h$ as the bandwidth, $(x_j, y_j)$ is the
rotated feature corresponding to $(b_j, d_j)$ from the persistence diagram, with $x_j =
(b_j+d_j)/2$ and $y_j = (d_j-b_j)/2$, and the $  \frac{y_j}{K_h(0)}$ ensures that
$\widetilde{\Lambda}_j$ goes through $(x_j, y_j)$.
For example, the Epanechnikov kernel would have $\widetilde{\Lambda}_j(t; h) =
(y_j) (1 - |\frac{t - x_j}{h}|^2)$ for $ | \frac{t - x_j}{h} | \leq 1$.
Using the new definitions of $\widetilde{\Lambda}_j$, the generalized landscapes are defined in the same manner as landscapes:
\begin{equation}
\widetilde{\mathbb{F}}_{k}(\PD;t) = \underset{j=1,\cdots, |\PD|}{\sf kmax} \{\widetilde{\Lambda}_j(t; h)\}, \label{eq:glandscape}.
\end{equation}

\begin{figure}
    \centering
\includegraphics[width=\textwidth]{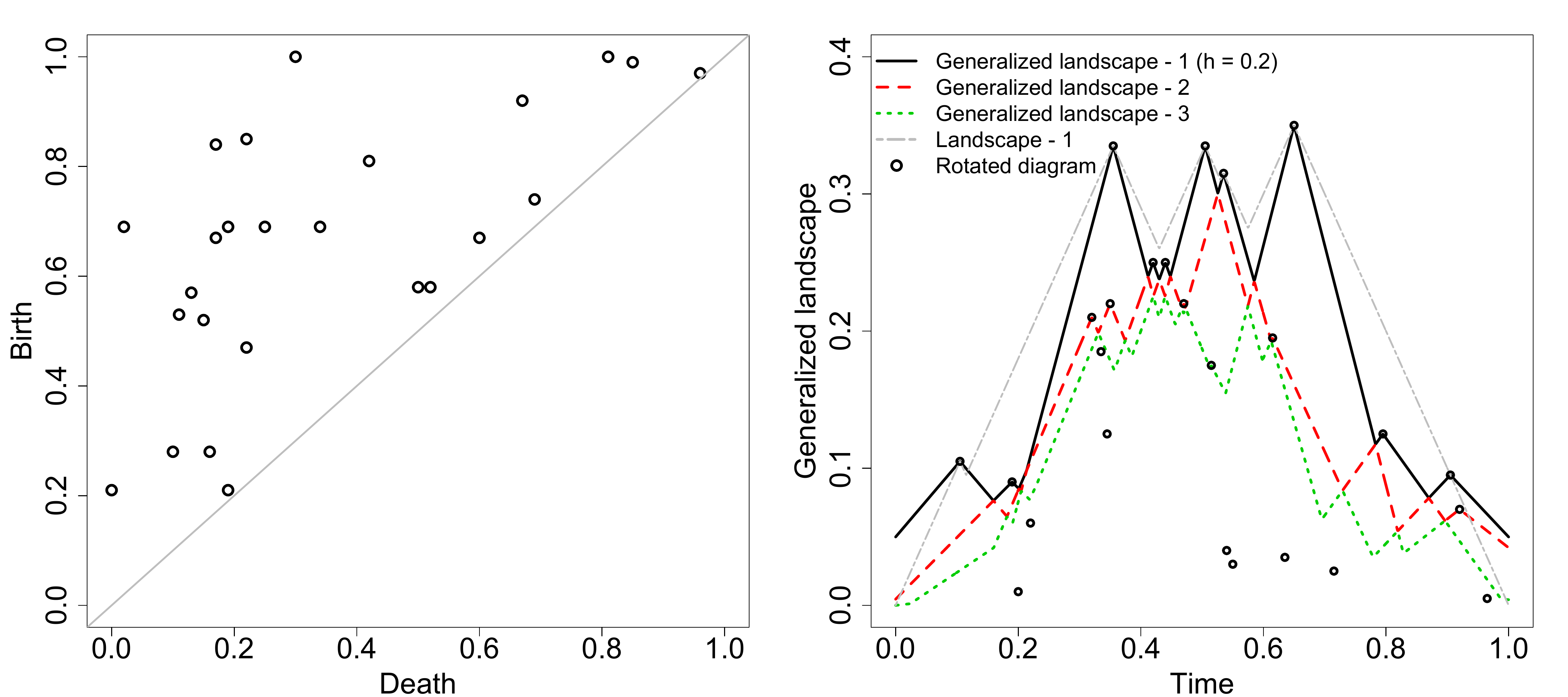}
    \caption{Generalized persistence landscape functions. The left plot is the persistence diagram used to produce the functional summaries in the right plot.  In the right plot, the black circles are the rotated persistence diagram features, the solid black line is the first generalized landscape using the triangle kernel with a bandwidth (half the triangle base) of $h = 0.2$, the second and third generalized landscapes are plotted in red dashes and green dots, respectively.  As a comparison, the first landscape function is plotted in gray as lines and dots. }
    \label{fig:gen_land}
\end{figure}

Illustrations of several generalized landscape functions are displayed in \figref{gen_land} using the triangle kernel.
As with general nonparametric smoothing methods, a smaller bandwidth results in a rougher function and a larger bandwidth results in a smoother function.  In Sections~\ref{subsec:gleasonExper} and \ref{sec:fibrin}, we consider both generalized landscapes and original landscapes in a classification problem and in two sample hypothesis tests.  Although the generalized landscapes and the landscape functions ultimately contain the same information, a benefit of using the generalized landscapes is that more features of the persistence diagram can be isolated to fewer function layers.

\section{Methodology} \label{sec:methods}
All functional summaries defined in the previous section map a persistence diagram to a function.
The problem of analyzing random persistence diagrams
becomes the problem of analyzing random functions,
which is a topic that has been studied for decades in statistics
\citep{van2000asymptotic}.
Next, we describe some analysis methods that can be done using the statistical
model characterized in~\eqref{eq:model}.

\subsection{Population quantities}
Using the statistical model of~\eqref{eq:model}, we can define
various population quantities associated to the functional summaries
such as the \emph{mean} functional summary
\citep{chazal2014stochastic,chen2015statistical,biscio2016accumulated}.
When averaging functional summaries of diagrams,
we obtain a sample mean functional summary.
This quantity can be used as an estimator of the population mean functional summary.
In more detail,
the population mean functional summary is a function
$$
\bar{F}(t) =  \mathbb{E}(F_i(t)),
$$
where the expectation $\E$ is with respect to the distribution $\mathcal{P}_F$.
The sample estimator is then the pointwise estimator
$$
\hat{F}(t) = \frac{1}{n}\sum_{i=1}^n F_i(t),
$$
and the convergence of $\hat{F}$ toward $\bar{F}$ can be studied
using notions of convergence of functions \citep{chazal2014stochastic,chen2015statistical}.

Let $\mathcal{B}_F$ be the set
of all possible functions formed by a given functional summary,
and let $\mathbb{T}$ be a compact set such that we are interested in the population mean functional summary $\bar{F}(t)$
within $t\in\mathbb{T}$.
To simplify the problem, we define $F(t)=0$ for all $t\notin \mathbb{T}$ for every $F\in\mathcal{B}_F$.
Throughout the paper, we assume that the functional summary is uniformly bounded
by a constant $\bar{U}<\infty$.
Namely,
\begin{equation}
\sup_{F\in\mathcal{B}_F}\sup_{x\in\mathbb{T}}|F(x)|\leq \bar{U}.
\label{eq::uniformbound}
\end{equation}

The following is a simple convergence theorem of the estimator $\hat{F}$, which shows that as long as the functional summary is continuous,
the sample mean function summary $\hat{F}$
uniformly converges to $\bar{F}$ almost surely, as expressed in the following proposition.
\begin{proposition}[Pointwise Convergence]
Assume Equation~\eqref{eq::uniformbound} holds.
If~$\mathcal{B}_F$ is equicontinuous, then
$$
\sup_{t\in\mathbb{T}}|\hat{F}(t)-\bar{F}(t)| \overset{a.s.}{\rightarrow}0.
$$
As a result, if there exists a constant $L>0$ such that any $F\in\mathcal{B}_F$ is $L$-Lipschitz, then
$$
\sup_{t\in\mathbb{T}}|\hat{F}(t)-\bar{F}(t)| \overset{a.s.}{\rightarrow}0.
$$
\label{prop::as}
\end{proposition}
The proof is presented in Appendix~\ref{sec::pf}.
Persistence intensity, persistence landscapes and generalized landscapes, and persistence scoring
all satisfy the condition in~\propref{:as} as long as the number of homological features
and the lifetime of the features are uniformly bounded.
Thus, averaging these functional summaries yields a consistent estimate of the corresponding
population mean functional summary.

The difference $\hat{F}(t) - \bar{F}(t)$ is tightly related to a normal distribution.
In what follows, we show that it converges to a normal distribution
in various ways.
Let $\mathcal{W} = \{\mathbb{F}_t:t\in \mathbb{T}\}$ such that $\mathbb{F}_t(D_i) = F_i(t)$
for each $i=1,\cdots, n$
be the mapping from the persistence diagram $D_i$ to a functional summary as
displayed in Equation~\eqref{eq:model}.
Let $Q$ be a probability measure over~$\mathcal{B}_F$
and define $\|f-g\|_{Q,2} = \sqrt{\int |f(t)-g(t)|^2dQ(t)}$
be the $L_2(Q)$ norm for functions
and
$N(\mathcal{W}, L_2(Q),\epsilon)$ be the $\epsilon$-covering number of $\mathcal{W}$.
\begin{proposition}
Let $\sigma^2(t)={\sf Var}(F_i(t))$ and $\sigma^2 = \int \sigma^2(t)dt$.
Assume Equation~\eqref{eq::uniformbound} holds, then
\begin{align*}
\sqrt{n}\left(\hat{F}(t) - \bar{F}(t)\right) &\overset{D}{\rightarrow}N(0,\sigma^2(t))\\
\sqrt{n}\int \left(\hat{F}(t) - \bar{F}(t)\right)dt &\overset{D}{\rightarrow}N(0,\sigma^2).
\end{align*}
Moreover, if
\begin{equation}
\int_{0}^1 \sqrt{\log \sup_{Q} N(\mathcal{W}, L_2(Q),\epsilon\bar{U})}d\epsilon<\infty,
\label{eq::entropy}
\end{equation}
where $\bar{U}$ is the upper bound of the functional summary
and the supremum is taken over all finitely discrete probability measures on the space of persistence diagrams
\footnote{
A finitely discrete probability measure $Q$ puts probability mass only finitely many
points in $\mathcal{B}_F$.},
then
$\sqrt{n}\left(\hat{F} - \bar{F}\right)$ converges in distribution to $\mathbb{B}$,
where $\mathbb{B}(t)$ is a Gaussian process over $t\in\mathbb{T}$ with a
covariance function
$$
{\sf Cov}(\mathbb{B}(t),\mathbb{B}(s))  = \mathbb{E}(F_i(t)F_i(s))-\bar{F}(t)\bar{F}(s)
$$
for $t,s\in\mathbb{T}$.
Futhermore,
if there exists a constant $L>0$ such that any $F\in\mathcal{B}_F$ is $L$-Lipschitz, then
the above three convergences hold.
\label{prop::normal}
\end{proposition}
The proof is presented in Appendix~\ref{sec::pf}.
Proposition \ref{prop::normal} presents the asymptotic normality
of the sample mean functional summary.
The assumptions are from the Donsker theorem; see, e.g., page 18 of \cite{kosorok2007introduction}.
The convergence toward a Gaussian process further implies
the convergence of the supremum,~i.e.,
$$
\sqrt{n}\sup_{t\in\mathbb{T}}|\hat{F}(t) - \bar{F}(t)| \overset{D}{\rightarrow} \sup_{t\in\mathbb{T}}|\mathbb{B}(t)|,
$$
by the continuous mapping theorem; see page 16 of \cite{kosorok2007introduction} for more detailed discussion.

The assumption \eqref{eq::entropy} is quite mild.
If the functional summary is $L$-Lipschitz,
then Equation~\eqref{eq::entropy} holds and thus $\sqrt{n}\left(\hat{F} - \bar{F}\right)$
converges to a Gaussian process.
A good news is that many functional summaries, such as the persistence landscapes, persistence silhouette,
persistence image, and persistence intensity are all $L$-Lipschitz functional summaries
under a very mild assumption (the number of features in the persistence diagrams is finite almost surely).
For the generalized landscape,
as long as the kernel function $K$ is Lipschitz (which is true for most of the common kernel functions),
the functional summary is also~Lipschitz.

\subsection{Confidence Bands}
Confidence bands provide a way to assess and visualize the uncertainty in the sample mean functional summary,
which can be constructed using a bootstrapping procedure \citep{chazal2014stochastic}.
Specifically, given a confidence level $\alpha$, the bootstrap can be employed to find a fixed bandwidth upper envelope function $U_{1-\alpha}(t)$
and a lower envelope function $L_{1-\alpha}(t)$
such that
\begin{align*}
P(L_{1-\alpha}(t) \leq \bar{F}(t) \leq U_{1-\alpha}(t)&\,\,\,\forall t\in[a,b])\\
&= 1-\alpha+o(1),
\end{align*}
for some regions $[a,b]$.  See Proposition~\ref{prop::CB} below for more details.

The following are the details of the construction of a confidence band.
Note that we assume the functional summary is a univariate function for simplicity; one can easily generalize
the following for multivariate functions.
\begin{enumerate}
\item {\bf The initial estimate.} First, compute the sample mean functional summary,
$
\hat{F}(t) = \frac{1}{n}\sum_{i=1}^n F_i(t),
$
for $t\in [a,b]$, a given interval.
\item {\bf The bootstrap procedure.} Sample diagrams with replacement and compute the corresponding
functional summaries and the sample mean functional summary,
denoted as $\hat{F}^*(t)$.
Namely, generate $\PD^*_1,\cdots,\PD^*_n$ by sampling randomly (with replacement) from $\PD_1,\cdots,\PD_n$ in a way such that
each diagram has an equal probability $(1/n)$ of being selected.
Then compute the the corresponding functional summaries $F^*_1,\cdots,F^*_n$, and calculate the sample mean functional summary, $\hat{F}^*(t)$.
\item {\bf Replication.} Repeat the bootstrap procedure $B$ times, leading to $B$ bootstrap realizations of the sample mean functional summary,
denoted as
$$
\hat{F}^{*(1)}(t),\cdots, \hat{F}^{*(B)}(t).
$$
\item {\bf Width of band.} For a given significance level $\alpha$, choose
\begin{equation*}
\hat{t}_{1-\alpha} = \hat{G}^{-1}(1-\alpha),\quad \hat{G}(s) = \frac{1}{B}\sum_{j=1}^B I\left(\sup_{t}\|\hat{F}^{*(j)}(t)-\hat{F}(t)\|\leq s\right),
\end{equation*}
where $\hat{t}_{1-\alpha}$ is the $1-\alpha$ quantile of the $L_\infty$-distance between the bootstrap realizations and the initial estimate.

\item {\bf Output.} The upper and lower bound of the confidence band is
$$
U_{1-\alpha}(t) = \hat{F}(t) + \hat{t}_{1-\alpha}, \quad L_{1-\alpha}(t) = \hat{F}(t) - \hat{t}_{1-\alpha}.
$$

\end{enumerate}

The following proposition shows that the confidence band is consistent under mild assumptions.
\begin{proposition}[Functional Bands]
If the assumptions in Proposition \ref{prop::normal}, including Equation~\eqref{eq::entropy}, hold,
then
\begin{align*}
P\large(L_{1-\alpha}(t)\leq \bar{F}(t)\leq U_{1-\alpha}(t)\,\,\forall t\in\mathbb{T}\large) = 1-\alpha+o(1).
\end{align*}
\label{prop::CB}
\end{proposition}
The proof is presented in Appendix~\ref{sec::pf}.
Proposition \ref{prop::CB} shows that the confidence band is asymptotically valid.
Note that if a slightly stronger assumption is~made,
\begin{equation}
\sup_{Q} N(\mathcal{W}, L_2(Q),\epsilon\bar{U}) \leq A\left(\frac{1}{\epsilon}\right)^\nu,
\label{eq::VC}
\end{equation}
for some constants $A,\nu>0$,
then
we can replace the $o(1)$ in proposition \ref{prop::CB}
by $O(n^{-1/8})$; see the derivation of \cite{chazal2014stochastic}.

The above procedure provides a fixed bandwidth confidence band.
A variable-bandwidth confidence band can also be constructed
using a simple modification in Step 4 and 5.
First, compute a variance estimator of the functional summaries
\begin{equation}
\hat{\sigma}^2(t) = \frac{1}{n}\sum_{i=1}^n\left(F_i(t)-\hat{F}(t)\right)^2.
\label{eq::var}
\end{equation}
Then, in Step 4, choose
\begin{equation*}
\hat{s}^\dagger_{1-\alpha} = \hat{G}_\dagger^{-1}(1-\alpha),\quad \hat{G}_\dagger(s) = \frac{1}{B}\sum_{j=1}^B I\left(\sup_{t}\left\|\frac{\hat{F}^{*(j)}(t)-\hat{F}(t)}{\hat{\sigma}(t)}\right\|\leq s\right).
\end{equation*}
And, in Step 5, we construct the band via
\begin{equation*}
U^\dagger_{1-\alpha}(t) = \hat{F}(t) + \hat{s}^\dagger_{1-\alpha}\cdot\hat{\sigma}(t), \quad L^\dagger_{1-\alpha}(t) =  \hat{F}(t) - \hat{s}^\dagger_{1-\alpha}\cdot\hat{\sigma}(t).
\end{equation*}
Using a similar derivation as \cite{chazal2014stochastic},
one can prove that such a confidence band is also valid

Note that both equations \eqref{eq::entropy} and \eqref{eq::VC}
hold if the functional summaries are $L$-Lipschitz.
Therefore, we have the following result for the generalized landscapes.

\begin{corollary}
Assume that $\mathbb{T}$ is compact
and the number of topological features of a persistence diagram is bounded by a constant
almost surely.
If the functional summaries are
constructed by a generalized persistence landscape
with a Lipschitz kernel function and a fixed $h>0$,
then the conclusions in proposition \ref{prop::as}, \ref{prop::normal},
and \ref{prop::CB}
are true.
\label{cor::generalize}
\end{corollary}
Corollary \ref{cor::generalize} is a direct result from the three propositions
when the kernel function is Lipschitz.
Most common kernel functions such as the triangle kernel and the Gaussian kernel
satisfy this condition.
Thus, this corollary implies that the generalized landscape function is a stable
functional summary for data analysis.

\subsection{Prediction Bands}
A sample mean functional summary $\hat{F}$
can be used to predict the
outcome of a future persistence functional summary.
Let $d:\mathcal{F}\times \mathcal{F}\mapsto \mathbb{R}$ be a metric for functional summaries.
For every functional summary, say $F_i$, we first compute the residual,
$
e_i = d(F_i,\hat{F}),
$
and then pick $\hat{q}_\gamma$ to be
the $\gamma$-quantile of $e_1,\cdots,e_n$.
Then a $\gamma$-prediction set is
\begin{equation}
\hat{\mathcal{F}}_\gamma = \{F: d(F,\hat{F})\leq \hat{q}_\gamma\}.
\label{eq::pred}
\end{equation}

The following Proposition \ref{prop::pred} proves that the prediction set in
Equation~\eqref{eq::pred}
is a valid prediction set in our setting.
\begin{proposition}
Let $d$ be a metric for functional summaries such that
$$
d(\hat{F}, \bar{F}) \overset{P}{\rightarrow} 0.
$$
Moreover, assume that the function
$$
Q(t) = P(d(F_i,\bar{F})<t)
$$
has a finite derivative bounded away from $0$ at an open neighborhood containing $t = q_\gamma$,
where $q_\gamma$ solves $Q(q_\gamma)=\gamma$.
Let $\hat{\mathcal{F}}_\gamma$ be as defined in Equation \eqref{eq::pred}.
Then
$$
P(F_{\sf new}\subset \hat{\mathcal{F}}_\gamma|F_1,\cdots,F_n) =\gamma+O_P\left(\left(\frac{1}{\sqrt{n}}\right)+d(\hat{F},\bar{F})\right).
$$
\label{prop::pred}
\end{proposition}
The proof is presented in Appendix~\ref{sec::pf}.
Note that the assumption $d(\hat{F}, \bar{F}) \overset{P}{\rightarrow} 0$
is very mild requiring that our estimator is consistent under the metric $d$.
The second assumption, the continuity of~$Q(t)$ at $t=q_\gamma$,
is also very weak; if the random variable $d(F_i,\bar{F})$ has a density function
that takes non-zero value around $t=q_\gamma$,
then this assumption holds.
Just as other propositions,
Proposition \ref{prop::pred} also applies to the generalized landscape
when the kernel function
is~Lipschitz.

There are many possible metrics $d$, and common choices are
\begin{equation}
\begin{aligned}
d_{p,\omega}(F, G) &= \left(\int \left(\frac{|F(t)-G(t)|}{\omega(t)}\right)^p dt\right)^{\frac{1}{p}}\\
d_{\infty,\omega}(F,G)& = \sup_t\left|\frac{F(t)-G(t)}{\omega(t)}\right|
\end{aligned}
\label{eq::metrics}
\end{equation}
where $\omega(t)>0$ is a weight function and $p$ is a positive number.
When $\omega(t) = 1$, $d_{p,\omega}$ becomes the $L_p$-metric for functions
and we often write $ d_p=d_{p,\omega} $ for simplicity.
$d_1$ and $d_2$ are popular choice in data analysis.
However,
the prediction set $\hat{\mathcal{F}}_\gamma$ from $d_1$ or $d_2$ (or any other $d_q$ for $0<q<\infty$) is hard to visualize.
The $L_\infty$ metric $d_\infty$ leads to a prediction set that is easy to visualize,
but $d_\infty$ can be too sensitive to small perturbations.

To obtain a stable metric with a simple visualization property,
we consider the metric $d_{\infty,\omega}$ with $\omega(t) = \hat{\sigma}(t) =\sqrt{\hat{\sigma}^2(t)}$,
the estimated standard deviation of $F_1(t), \ldots, F_n(t)$ from Equation \eqref{eq::var}.
This leads to a variable-width prediction band.
The prediction band can be constructed by first computing
$
e_i = d_{\infty, \hat{\sigma}}(F_i,\hat{F})
$
for each $i=1,\cdots,n$.
Given a prediction level $\gamma>0$,
let $$
\hat{q}_\gamma = \gamma\mbox{-quantile of }\{e_1,\cdots,e_n\}.
$$
Then the $\gamma$-prediction band is
\begin{equation}
\left\{f(t): \tilde{L}_\gamma(t)\leq f(t)\leq \tilde{U}_\gamma(t), t\in  \mathbb{T}\right\},
\label{eq::pred::w1}
\end{equation}
where
\begin{equation}
\tilde{U}_\gamma(t) = \hat{F}(t) + \hat{q}_\gamma\cdot\hat{\sigma}(t) \text{ and } \tilde{L}_\gamma(t) = \hat{F}(t) - \hat{q}_\gamma\cdot\hat{\sigma}(t).
\label{eq::pred::w2}
\end{equation}
Thus,
a simple way to visualize the prediction set
is to plot a band governed by the lower envelope $\tilde{L}_\gamma(t)$ and the upper envelope $\tilde{U}_\gamma(t)$.

\begin{remark}[Conformal prediction band]
The above prediction bands have $1-\alpha$ coverage asymptotically.
If we want to obtain a prediction band with exact $1-\alpha$ coverage,
we may use the conformal prediction approach \citep{vovk2005algorithmic,shafer2008tutorial, lei2015conformal}.
\cite{lei2015conformal}
proposed two methods of constructing a prediction band
for functional data combining the data splitting and conformal prediction, which can be used to construct an exact $1-\alpha$ prediction band
for functional summaries.
\end{remark}

\subsection{Two-Sample Test} \label{sec:two_sample}
Because functional summaries can be averaged,
two-sample tests of two groups of persistence diagrams can be carried out \citep{chen2015statistical,biscio2016accumulated}.
This scenario is common in biomedical applications where one set of diagrams comes from a control group and another set from a
treatment group.

There are many ways to carryout two-sample tests, but here we consider a permutation test approach.
Assume we observe two sets of diagrams
\begin{equation*}
\PD_{1,1},\cdots,\PD_{1,n} \sim \mathcal{P}_{F,1}, \quad \PD_{2,1},\cdots,\PD_{2,m} \sim \mathcal{P}_{F,2},
\end{equation*}
then the goal is to test the null hypothesis
\begin{equation}
H_0: \mathcal{P}_{F,1}=\mathcal{P}_{F,2}. \label{eq:two_sample}
\end{equation}
Namely, we want to see if there is evidence suggesting the two sets of diagrams were sampled from different populations.

Let $F_{\ell,i} = F(\PD_{\ell,i})$ be the functional summaries of the corresponding persistence diagrams for set $\ell$,
and let
$$
\hat{F}_{1}(t)=\frac{1}{n}\sum_{i=1}^n F_{1,i}(t),\quad\hat{F}_2(t)= \frac{1}{m}\sum_{i=1}^m F_{2,i}(t)
$$
denote the sample mean functional summary of each group.
To perform a permutation test, we first choose a metric of functions, such as
one from Equation \eqref{eq::metrics},
and then define
$$
T  = d(\hat{F}_1, \hat{F}_2)
$$
as the test statistic.
To compute the permutation p-value,
the functional summaries from both samples are combined and then randomly split into two groups,
one group with $n$ functions and the other with $m$ functions.
New sample mean functional summaries
of both groups are computed along with the test statistic using the new averages.
Assume the above procedure is repeated $B$ times, then $T^{*(1)},\cdots,T^{*(B)}$ realizations
of the test statistic are obtained.
The p-value is the proportion of $T^{*(j)}$ that are greater than or equal to the original test statistic $T$, i.e.,
$$
\mbox{permutation p-value}=\frac{1}{B}\sum_{j=1}^B I\left(T^{*(j)}\geq T\right).
$$
The idea of permutation test is that when $H_0$ is true, permuting the
diagrams or their corresponding functions does not change the distribution significantly.
A powerful feature of the permutation test is that it is test; namely, the significance level $\alpha$ can be controlled
at any level exactly \citep{wasserman2006all}.

\subsection{Classification}

Using the statistical model \eqref{eq:model}, the problem of classification of
persistence diagrams can be studied \citep{biscio2016accumulated}.
We consider the binary classification for simplicity, which can be generalized to multiple
classes.
The problem of classification is as follows.
Suppose a collection of diagrams with labels are observed as
$$
(\PD_1,Y_1),\cdots, (\PD_n, Y_n),
$$
where each $Y_i \in \{0,1\}$ denotes the class label of $i$-th diagram.
Statistical classification addresses how to predict the class label of a new persistence diagram, $\PD_{\sf new}$.
Using functional summaries, the goal is to predict the class label for $F_{\sf new} = F(\PD_{\sf new})$.
This can be solved by techniques from classifying functional data \citep{wang2015review}.

\begin{sloppypar}
Here, we provide a simple approach based on the $k$-nearest neighbor~(kNN) classifier.
First a metric $d$ is chosen, possibly from Equation~\eqref{eq::metrics},
and then a new functional summary $F_{\sf new}$, is used to compute the distances
$d(F_{1},F_{\sf new}),\cdots, d(F_{n},F_{\sf new})$.
Let $F_{(\ell)}$ denotes the $\ell$-th closest functional summary to $F_{\sf new}$
and $Y_{(\ell)}$ be its corresponding class label.
The kNN classifier is~then
$$
c_k(F_{\sf new}) =
\begin{cases}1,\quad &\mbox{if }\sum_{\ell=1}^k Y_{(\ell)} > \frac{k}{2} \\
0,\quad &\mbox{if }\sum_{\ell=1}^k Y_{(\ell)} \leq \frac{k}{2}.
\end{cases}
$$
Namely, if more than half of $F_{\sf new}$'s $k$ neighborhoods have a class label $1$,
the label of $\PD_{\sf new}$ is $1$;
otherwise the label of $\PD_{\sf new}$ is $0$.
In practice, one can choose~$k$ by minimizing the classification errors using
cross-validation.
\end{sloppypar}

\subsection{Clustering and Visualization}

There are two major approaches for clustering diagrams based on functional summaries.
The first approach is to directly cluster functional summaries using
clustering techniques from functional data analysis.
For instance, one can use functional k-means clustering \citep{wang2015review} or mode clustering \citep{ciollaro2014functional}
to separate functional summaries into clusters and partition diagrams accordingly.

The other approach is based on a pairwise distance matrix.
A metric is chosen, possibly from Equation~\eqref{eq::metrics},
and then a pairwise distance matrix of functional summaries is computed.
Based on the distance matrix, clustering can be carried out using
ideas similar to spectral clustering or hierarchical clustering to partition the functions into several clusters \citep{von2007tutorial, jacques2014functional}.

An extra advantage of the second approach is that
it automatically provides a way to visualize the distance/similarity between
diagrams.
Using the distance matrix, the classical multidimensional scaling (or other approaches) can be performed
to see how each diagram is related to one another \citep{chen2015statistical,biscio2016accumulated}.

\begin{remark}[Connection to Functional Data Analysis]
In all the above analyses,
a number of statistical approaches are available from functional data analysis \citep{ferraty2006nonparametric,ramsay2006functional, wang2015review}.
This is because the functional summaries of \eqref{eq:model}
map the persistence diagrams into functions that provide a link between
the two fields.
Because of this mapping,
one can apply the tools from functional data analysis
to analyze the persistence diagrams \citep{biscio2016accumulated}
and the current research along this direction has yielded many
fruitful results \citep{chazal2014stochastic, adams2015persistence, chen2015statistical,biscio2016accumulated}.
\end{remark}

\section{Experimental Evaluation} \label{sec:sim_study}

Next we consider the two applications introduced in Section~\ref{sec:intro}: the Gleason data and the fibrin network data.  These datasets are considered because they highlight two different types of persistence diagrams.  The Gleason data has an underlying loop structure with topological randomness due to sampling and to variability around the shape of the loop (indicating a spectrum from benign to cancerous) with a persistence diagram containing fewer features.
While the spatially complex fibrin data has a complicated persistence diagram with many features.  Classification and two-sample tests are carried out using functional summaries of the persistence diagrams to investigate their effectiveness.

\subsection{Simulated Gleason Data}\label{subsec:gleasonExper}

\begin{figure}
    \centering
    \begin{subfigure}{0.22\textwidth}
        \centering
        \includegraphics[height=1.1in]{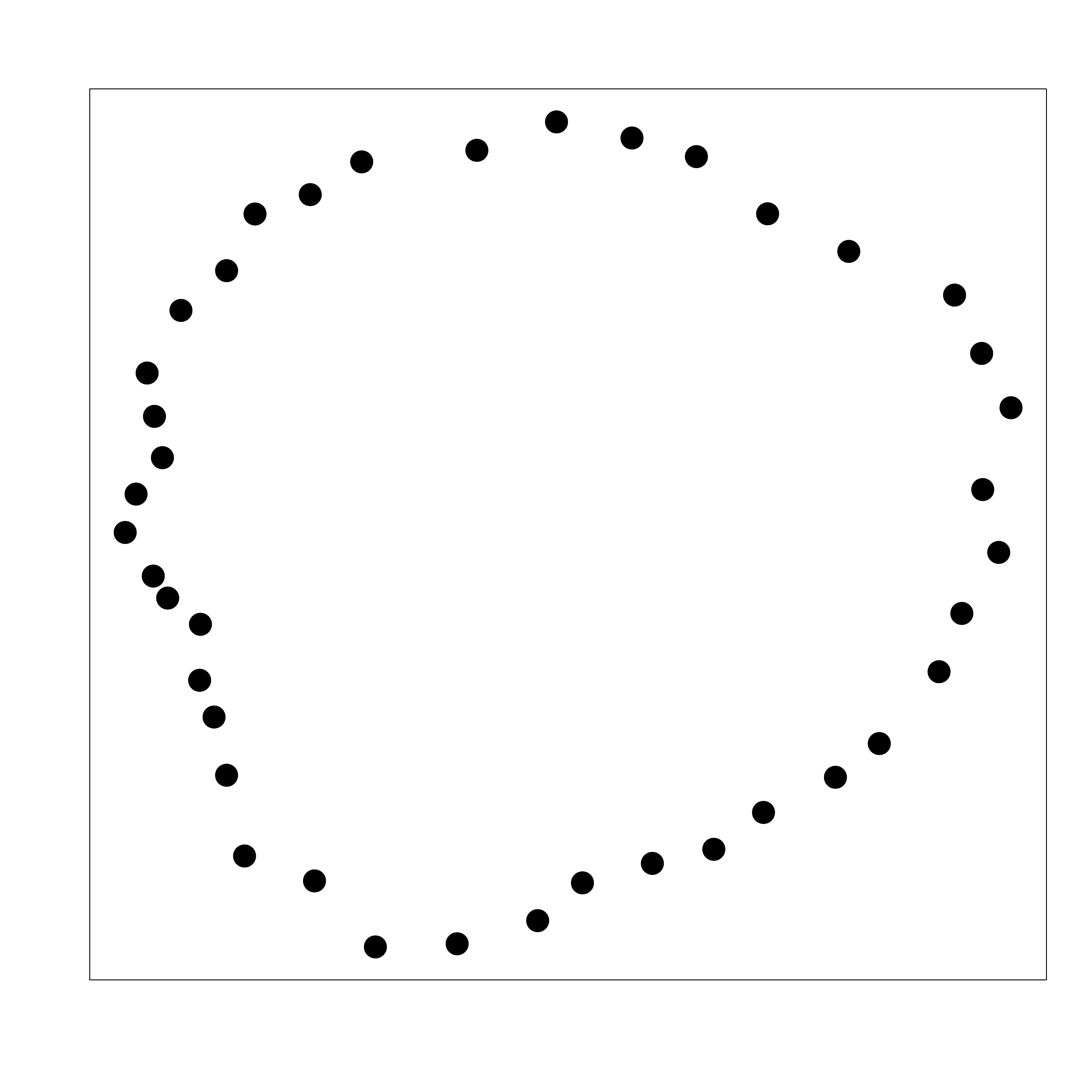}
        \caption{Type A}
    \end{subfigure}
    \begin{subfigure}{0.22\textwidth}
        \centering
        \includegraphics[height=1.1in]{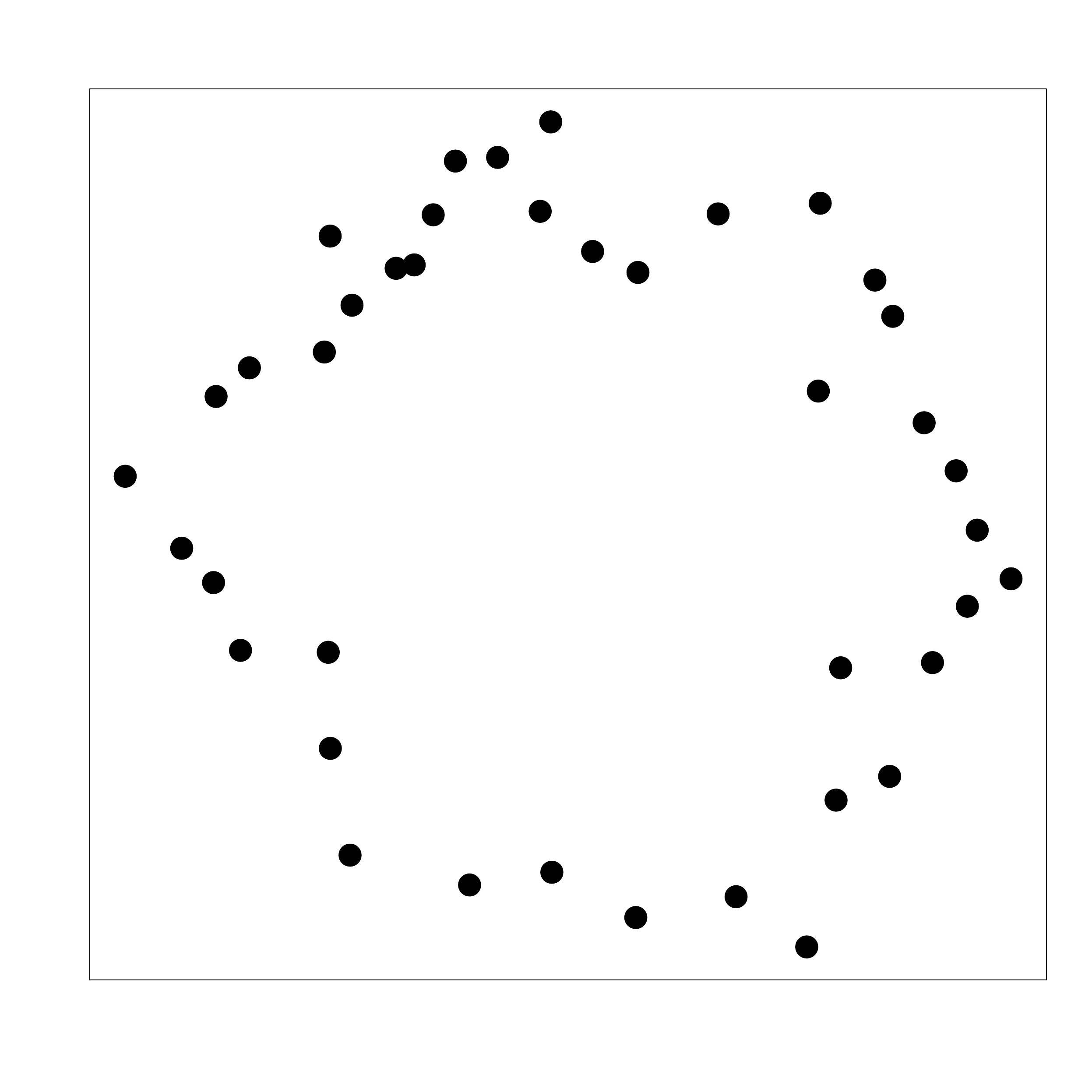}
        \caption{Type B}
    \end{subfigure}
    \begin{subfigure}{0.22\textwidth}
        \centering
        \includegraphics[height=1.1in]{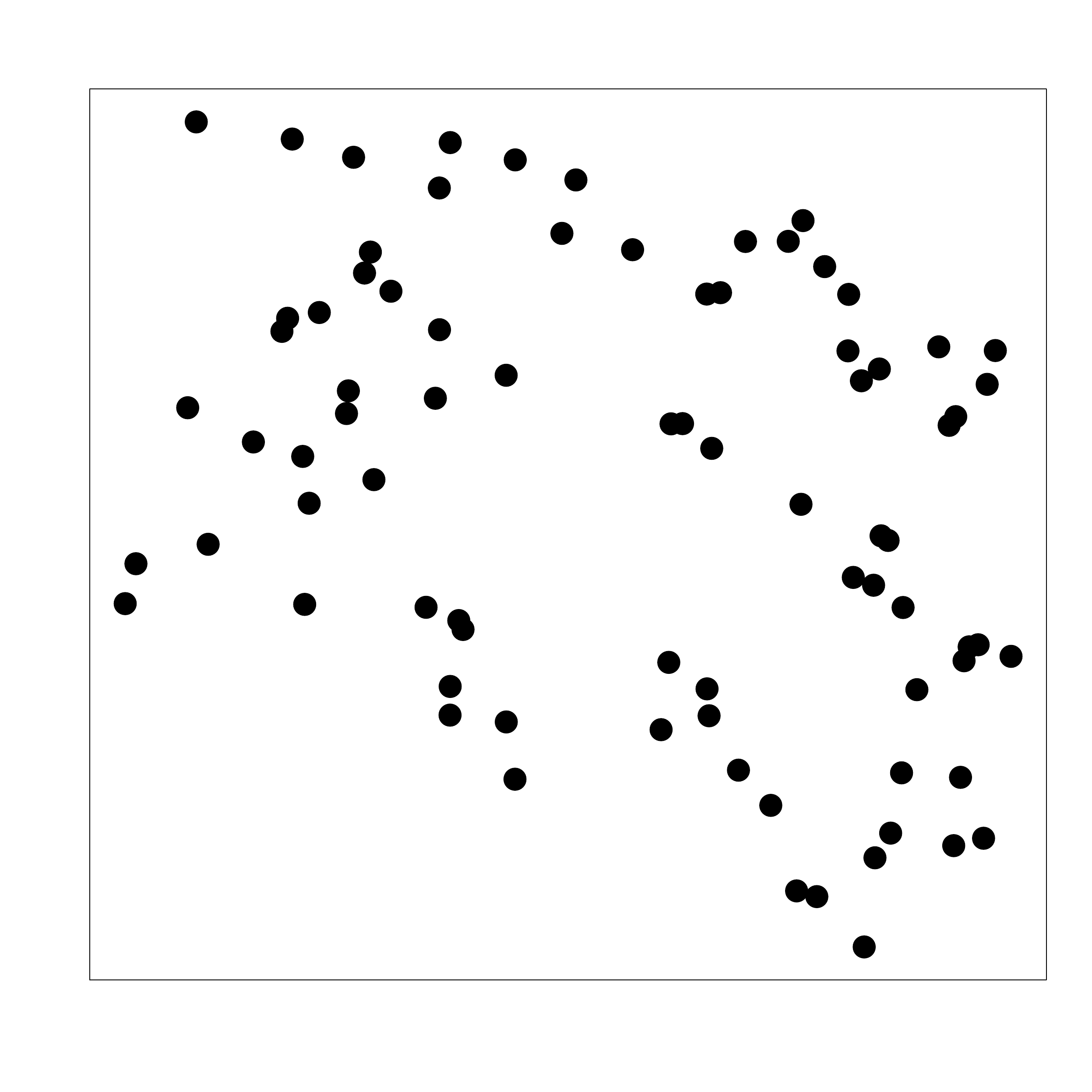}
        \caption{Type C}
    \end{subfigure}
    \begin{subfigure}{0.22\textwidth}
        \centering
        \includegraphics[height=1.1in]{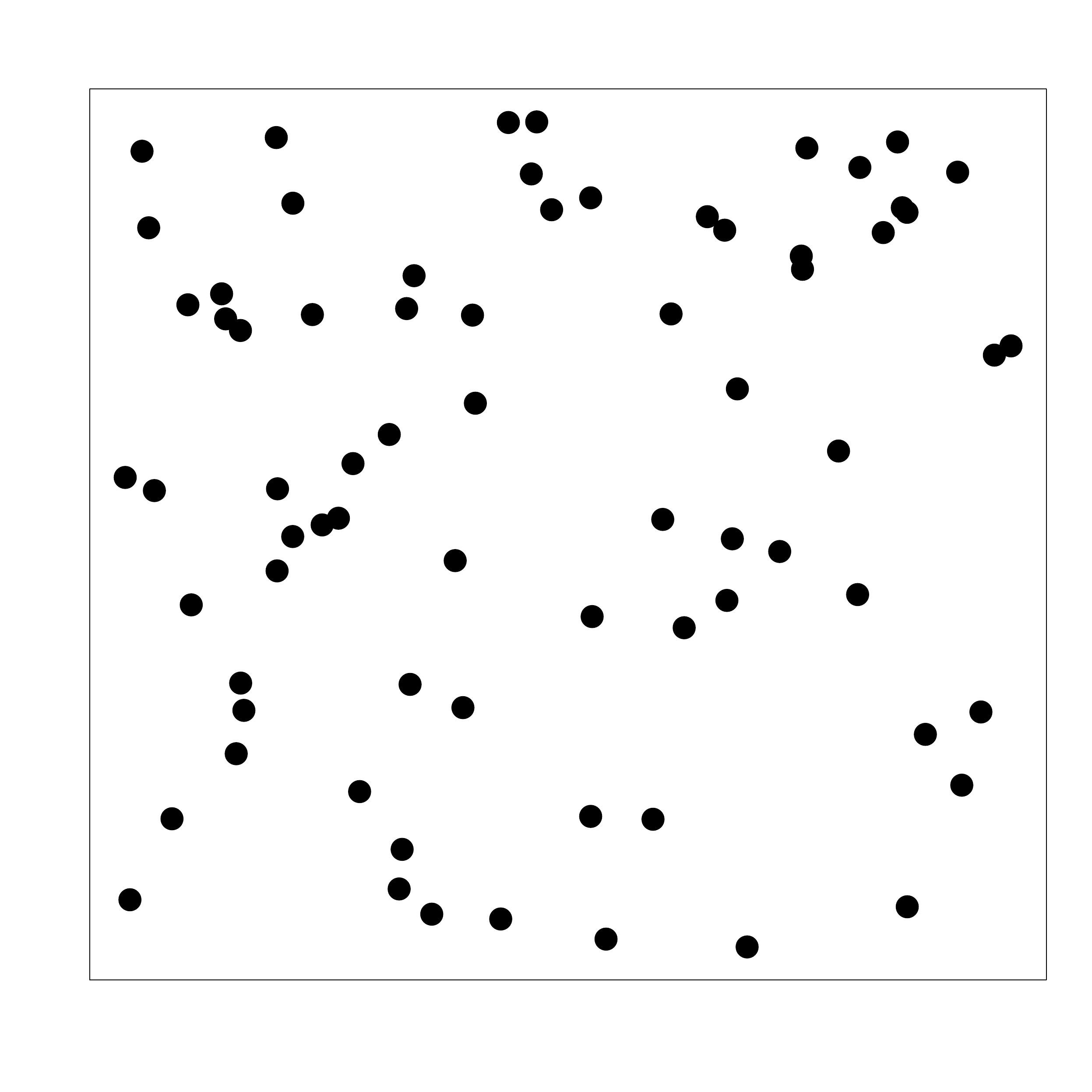}
        \caption{Type D}
    \end{subfigure}
    \caption{These glandular structures model typical glands found in prostate biopsy slides, with type A being
        the least cancerous, and type D being the most cancerous.
    }
    \label{fig:glands}
\end{figure}

Preliminary analysis has shown that clustering regions of interest (ROIs)  correlates with the
Gleason grades for purely-graded
regions \citep{lawson-spie2017,lawson-architectural}.  In an effort to
both better understand the progression of cancer and to curate a larger data set
with known / controlled grading, and because obtaining digital slides is both expensive and time-consuming, we are
developing a mock slide synthesizer~\citep{gland-gen}.  The data we study in this paper are
obtained from the gland synthesizer used in the mock slide synthesizer, where a \emph{benign} gland would be one
that is round, with a diameter about $80$-$100 \mu$m (similar to
glands found in Grade $3$ ROIs).  The \emph{unhealthy} state is one where the
gland and cribriform are indistinguishable, leaving what looks like a sheet of
cells (i.e., the nuclei appear to be uniformly distributed).  With a tuning
parameter, we create cells that range from \emph{benign} (Type A) to less and
less healthy (Types B and C, respectively) to \emph{unhealthy} (Type D); see
\figref{glands}. Note that we refer to these as Types A--D, rather than using the usual Gleason grading
scheme of, for example, grades 1--5, to emphasize that our data is computationally generated and the types have not been
verified by a pathologist to correspond to particular Gleason grades.

To illustrate how functional summaries can be useful in analyzing Gleason data,
we conduct a classification analysis with functional summaries.
Moreover, we compare the performance in terms of classification based on three functional summaries:
the silhouette functions, the landscape functions, and the generalized landscape functions with the triangle kernel.
Specifically, we performed $k$-nearest neighbors ($k$NN) classification using the $L_2$-metric
with $k$ being chosen by the leave-one-out cross-validation (LOOCV).
We created a training set consisting of $2000$ simulated glands with $500$ of each
type (A--D), and computed
persistence diagrams for each observation.
Then we generated a test set of 400 simulated glands, $100$ of each type,
and computed the corresponding persistence diagrams as well.

\begin{figure}
    \begin{subfigure}{0.45\textwidth}
            \includegraphics[height=2.5in]{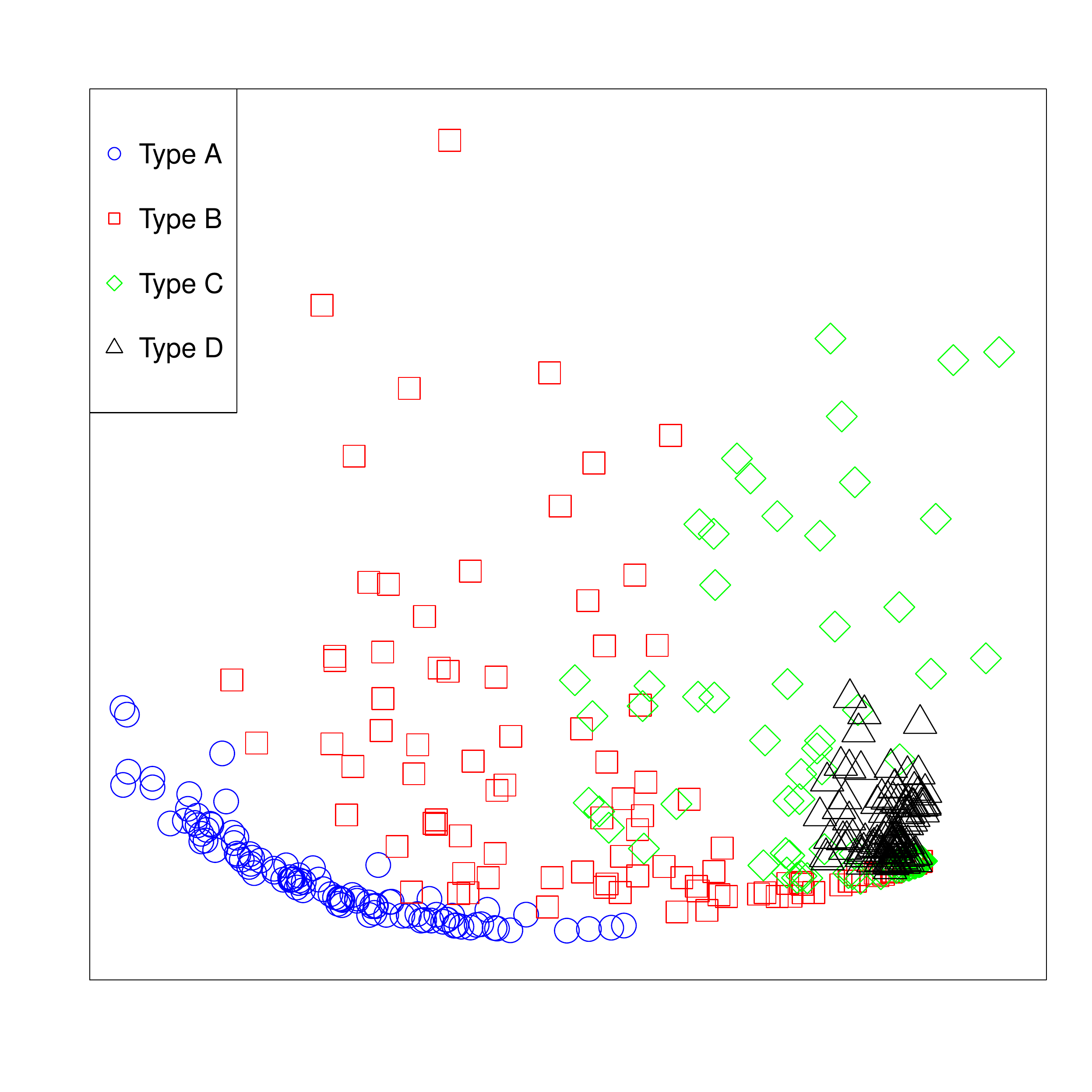}%
            \caption{MDS Plot}\label{mds}
    \end{subfigure}
    \quad \quad
    \begin{subfigure}{0.45\textwidth}
        \centering
        \vspace{.9in}
    \begin{tabular}{|c|c|c|c|c|}
            \hline Type &  A &  B &  C &  D \\ \hline
             A & 99 & 1 & 0 & 0 \\ \hline
             B & 1 & 86 & 13 & 0 \\ \hline
             C& 0 & 10 & 76 & 14 \\ \hline
             D & 0 & 0 & 8 & 92 \\ \hline
            \multicolumn{1}{c}{} & \multicolumn{1}{c}{} & \multicolumn{1}{c}{} & \multicolumn{1}{c}{}
                & \multicolumn{1}{c}{} \\
            \multicolumn{1}{c}{} & \multicolumn{1}{c}{} & \multicolumn{1}{c}{} & \multicolumn{1}{c}{}
                & \multicolumn{1}{c}{} \\
            \multicolumn{1}{c}{} & \multicolumn{1}{c}{} & \multicolumn{1}{c}{} & \multicolumn{1}{c}{}
                & \multicolumn{1}{c}{} \\
            \multicolumn{1}{c}{} & \multicolumn{1}{c}{} & \multicolumn{1}{c}{} & \multicolumn{1}{c}{}
                & \multicolumn{1}{c}{} \\
        \end{tabular}
        \vspace{.4in}
        \caption{Confusion Matrix}\label{confusion}
    \end{subfigure}%
        \caption{(left) A two-dimensional MDS plot of the test set using silhouette functions.
                (right)
Silhouette function classification test results.  Confusion matrix for the knn classification tests given as percentages using silhouette functions.
            The rows are the actual type, and the columns
            are the classified type.
                }
\end{figure}

Our first analysis was done using the silhouette
function.
For each persistence diagram, a silhouette
function was computed with weights equal to the lifetime of each feature.
We apply LOOCV to the training set to choose the number $k$ that minimizes the classification error on the training set.
With this choice of $k$, we use the entire training set to construct a $k$NN classifier.
To evaluate the performance of this classifier, we use the test set,
which leads to
an $11.75\%$ overall test classification error, with 47 glands misclassified.
Figure \ref{confusion} displays the confusion matrix for the silhouette functions.
To demonstrate an approach for visualizing functional summaries,
we used multidimensional scaling (MDS; see, .e.g, \citealt{friedman2001elements}) on the silhouette functions
to obtain a two-dimensional visualization of the true classes, as seen in
Figure~\ref{mds}.

\begin{sloppypar}
This exercise was repeated for landscape functions and generalized landscape functions with the triangle kernel and bandwidths of $h = 0.01, 0.025, 0.05, 0.10,$ and $0.25$.
Since the landscape approach (generalized landscape) contains several landscape functions,
a decision has to be made regarding the number functions to use for each setting.
We carried out the simulation study using  $1:j$ functions for $j = 1, \ldots, 6$
(i.e. only using the first function, using only the first and second function, to using the first through sixth functions).
For each function type, bandwidth, and number of functions,
we apply LOOCV to the training set to choose the tuning parameter $k$,
then construct the $k$NN classifier using the training set,
and evaluate the accuracy using the test set, as was done for the silhouette functions.
Note that in the case of generalized landscape function, we also choose the smoothing bandwidth using LOOCV.
The results for the landscape functions are displayed in Table~\ref{confusion2},
and the results for the best generalized landscape setting (which was the setting with a bandwidth of $h = 0.25$) are displayed in Table~\ref{confusion3}.
The best setting for the landscape functions was the case using landscapes $1$
    through~$6$ with a total number of 57 misclassifications resulting in an overall test classification error rate of 14.25\%.
The best setting for the generalized landscape functions with $h = 0.25$ were
    the cases using generalized landscapes $1$ through~$5$, or $1$ through $6$, with a total number of 39 misclassifications resulting in an overall classification error rate of 9.75\%.

\end{sloppypar}

Between the silhouette functions, landscape functions, and generalized landscape
functions considered in this study, the best generalized landscape functions had
the best overall performance in terms of test misclassification error.
Most of the errors in classification
occurred with gland types B and C where true gland type B was sometimes
misclassified as type C, and true gland type C was sometimes misclassified as
type B or D.
The generalized landscape functions are able to include more
details in fewer functions, which may explain why they performed better than the
landscapes functions.
Silhouette functions contain more layers of information
than the landscape functions considered, though the information is averaged
based on the lifetime of the features, which may explain their resulting
performance between that of the landscape and silhouette functions.

\begin{table}[htbp]
        \begin{tabular}{|c|c|c|c|c|}
       \multicolumn{5}{c}{(a) Function orders 1} \\
            \hline Type &  A &  B &  C &  D \\ \hline
A &	98	&	2	&	0	&	0	\\ \hline
B &	1	&	82	&	17	&	0	\\ \hline
C &	0	&	17	&	69	&	14	\\ \hline
D &	0	&	1	&	14	&	85	\\ \hline
   \end{tabular} \quad
           \begin{tabular}{|c|c|c|c|c|}
                  \multicolumn{5}{c}{(b) Function orders 1:2} \\
            \hline Type &  A &  B &  C &  D \\ \hline
A &	98	&	2	&	0	&	0	\\ \hline
B &	1	&	86	&	13	&	0	\\ \hline
C &	0	&	16	&	71	&	13	\\ \hline
D &	0	&	0	&	13	&	87	\\ \hline
   \end{tabular} \\ \bigskip
         \begin{tabular}{|c|c|c|c|c|}
                \multicolumn{5}{c}{(c) Function orders 1:3} \\
            \hline Type &  A &  B &  C &  D \\ \hline
A &	98	&	2	&	0	&	0	\\ \hline
B &	1	&	84	&	15	&	0	\\ \hline
C &	0	&	16	&	70	&	14	\\ \hline
D &	0	&	0	&	12	&	88	\\ \hline
   \end{tabular} \quad
           \begin{tabular}{|c|c|c|c|c|}
                  \multicolumn{5}{c}{(d) Function orders 1:4} \\
            \hline Type &  A &  B &  C &  D \\ \hline
A &	98	&	2	&	0	&	0	\\ \hline
B &	1	&	82	&	17	&	0	\\ \hline
C &	0	&	16	&	70	&	14	\\ \hline
D &	0	&	0	&	12	&	88	\\ \hline
   \end{tabular}     \\ \bigskip
           \begin{tabular}{|c|c|c|c|c|}
                  \multicolumn{5}{c}{(e) Function orders 1:5} \\
            \hline Type &  A &  B &  C &  D \\ \hline
A &	99	&	1	&	0	&	0	\\ \hline
B &	1	&	83	&	16	&	0	\\ \hline
C &	0	&	16	&	70	&	14	\\ \hline
D &	0	&	0	&	13	&	87	\\ \hline
   \end{tabular} \quad
           \begin{tabular}{|c|c|c|c|c|}
                  \multicolumn{5}{c}{(f) Function orders 1:6} \\
            \hline Type &  A &  B &  C &  D \\ \hline
A &	99	&	1	&	0	&	0	\\ \hline
B &	1	&	85	&	14	&	0	\\ \hline
C &	0	&	16	&	70	&	14	\\ \hline
D &	0	&	0	&	11	&	89	\\ \hline
   \end{tabular}
\caption{Landscape function gland classification test results.  Confusion matrix
    for the $k$nn classification tests given as percentages using landscape
    functions.  (a)--(f) are the results using function orders 1, 1:2, 1:3, 1:4, 1:5, and 1:6, respectively.
The number of nearest neighbors, $k$, was selected using leave-one-out cross-validation for each setting using the training data, and then the test data was used with the optimal $k$ for each setting to obtain the correct classification percentages displayed in these tables.
The rows are the actual gland type, and the columns are the classified gland type.
    }\label{confusion2}%
\end{table}

\begin{table}[htbp]
        \begin{tabular}{|c|c|c|c|c|}
       \multicolumn{5}{c}{(a) Function orders 1} \\
            \hline Type &  A &  B &  C &  D \\ \hline
A &	100	&	0	&	0	&	0	\\ \hline
B &	1	&	89	&	10	&	0	\\ \hline
C &	0	&	10	&	77	&	13	\\ \hline
D &	0	&	2	&	8	&	90	\\ \hline
   \end{tabular} \quad
           \begin{tabular}{|c|c|c|c|c|}
                  \multicolumn{5}{c}{(b) Function orders 1:2} \\
            \hline Type &  A &  B &  C &  D \\ \hline
A &	100	&	0	&	0	&	0	\\ \hline
B &	1	&	89	&	10	&	0	\\ \hline
C &	0	&	12	&	71	&	17	\\ \hline
D &	0	&	0	&	5	&	95	\\ \hline
   \end{tabular} \\ \bigskip
         \begin{tabular}{|c|c|c|c|c|}
                \multicolumn{5}{c}{(c) Function orders 1:3} \\
            \hline Type &  A &  B &  C &  D \\ \hline
A &	100	&	0	&	0	&	0	\\ \hline
B &	1	&	89	&	10	&	0	\\ \hline
C &	0	&	15	&	72	&	13	\\ \hline
D &	0	&	0	&	7	&	93	\\ \hline
   \end{tabular} \quad
           \begin{tabular}{|c|c|c|c|c|}
                  \multicolumn{5}{c}{(d) Function orders 1:4} \\
            \hline Type &  A &  B &  C &  D \\ \hline
A &	100	&	0	&	0	&	0	\\ \hline
B &	1	&	91	&	8	&	0	\\ \hline
C &	0	&	14	&	76	&	10	\\ \hline
D &	0	&	1	&	6	&	93	\\ \hline
   \end{tabular}     \\ \bigskip
           \begin{tabular}{|c|c|c|c|c|}
                  \multicolumn{5}{c}{(e) Function orders 1:5} \\
            \hline Type &  A &  B &  C &  D \\ \hline
A &	100	&	0	&	0	&	0	\\ \hline
B &	1	&	91	&	8	&	0	\\ \hline
C &	0	&	14	&	78	&	8	\\ \hline
D &	0	&	1	&	7	&	92	\\ \hline
   \end{tabular} \quad
           \begin{tabular}{|c|c|c|c|c|}
                  \multicolumn{5}{c}{(f) Function orders 1:6} \\
            \hline Type &  A &  B &  C &  D \\ \hline
A &	100	&	0	&	0	&	0	\\ \hline
B &	1	&	91	&	8	&	0	\\ \hline
C &	0	&	14	&	78	&	8	\\ \hline
D &	0	&	1	&	7	&	92	\\ \hline
   \end{tabular}
        \caption{Generalized landscape function gland classification test
        results. Confusion matrix for the $k$nn classification tests given as
        percentages using generalized landscape functions with triangle kernels
        and a bandwidth of 0.25.  (a)--(f) are the results using function orders 1, 1:2, 1:3, 1:4, 1:5, and 1:6, respectively.
The number or nearest neighbors, $k$, was selected using leave-one-out cross-validation for each setting using the training data, and then the test data was used with the optimal $k$ for each setting to obtain the correct classification percentages displayed in these tables.
The rows are the actual gland type, and the columns are the classified gland type.}\label{confusion3}%
\end{table}

\subsection{Fibrin Data} \label{sec:fibrin}
\begin{figure}
    \centering
     \begin{subfigure}{0.49\textwidth}
        \centering
        \includegraphics[height=2in]{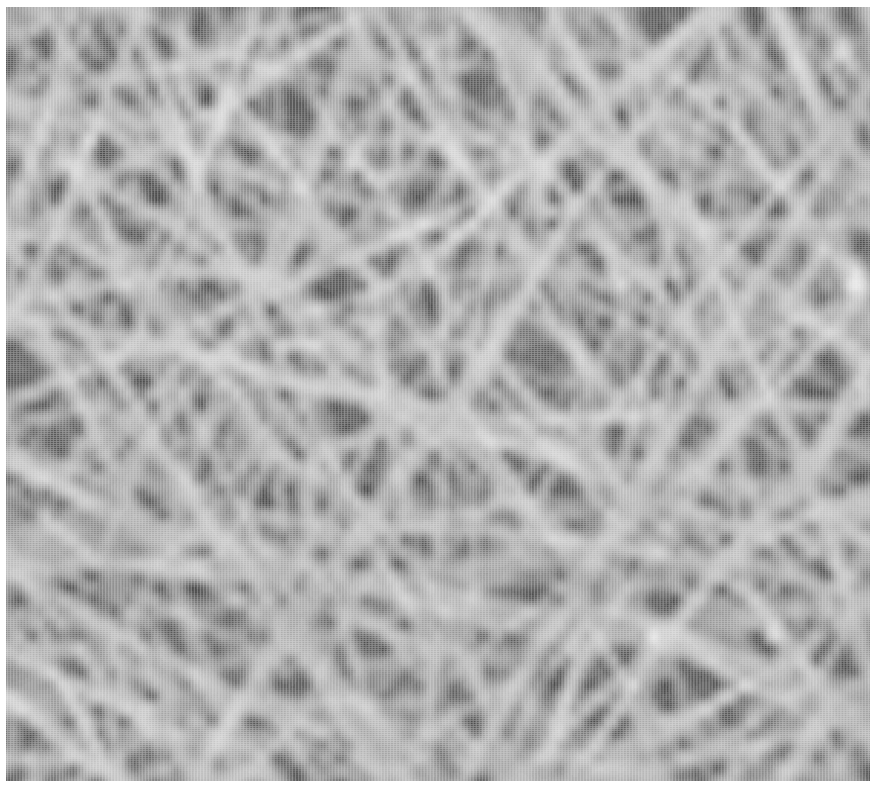}
        \caption{Modeled human fibrin}\label{subfig:modeled_human}
    \end{subfigure}
    \begin{subfigure}{0.49\textwidth}
        \centering
        \includegraphics[height=2in]{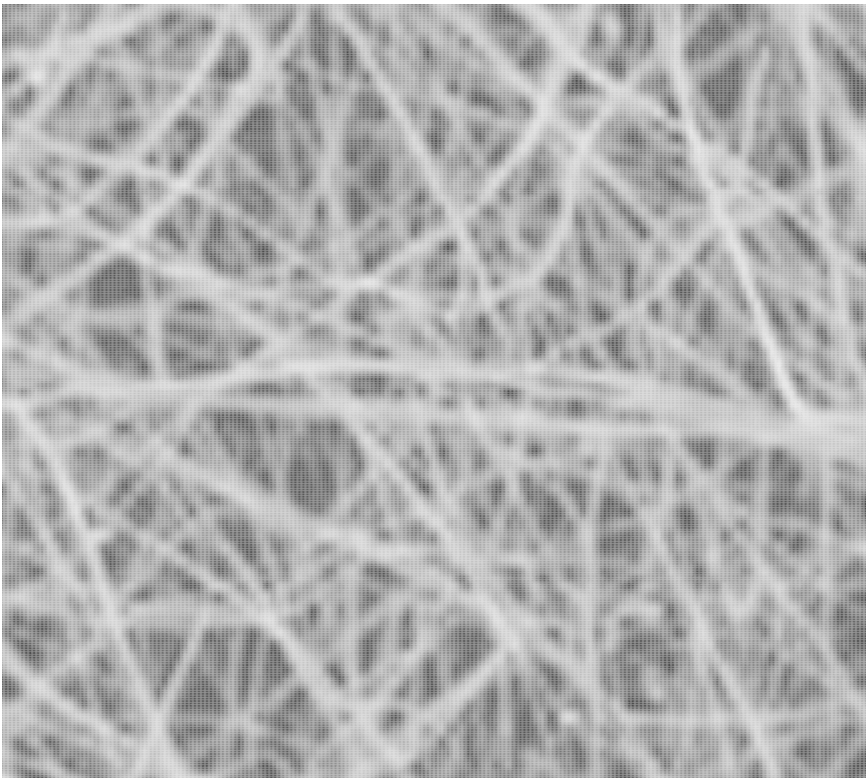}
        \caption{Modeled monkey fibrin}\label{subfig:modeled_monkey}
    \end{subfigure}
    \caption{Modeled human fibrin network (left) and monkey fibrin network
    (right); original images are from \cite{Pretorius:2009aa}.  The modeling
    step uses local quadratic regression with an adaptive bandwidth that
    includes the 0.1\% of the nearest neighbors of the point of
    interest.} \label{fig:human_monkey}
\end{figure}

\begin{figure}
    \centering
     \begin{subfigure}{0.49\textwidth}
        \centering
        \includegraphics[height=2in]{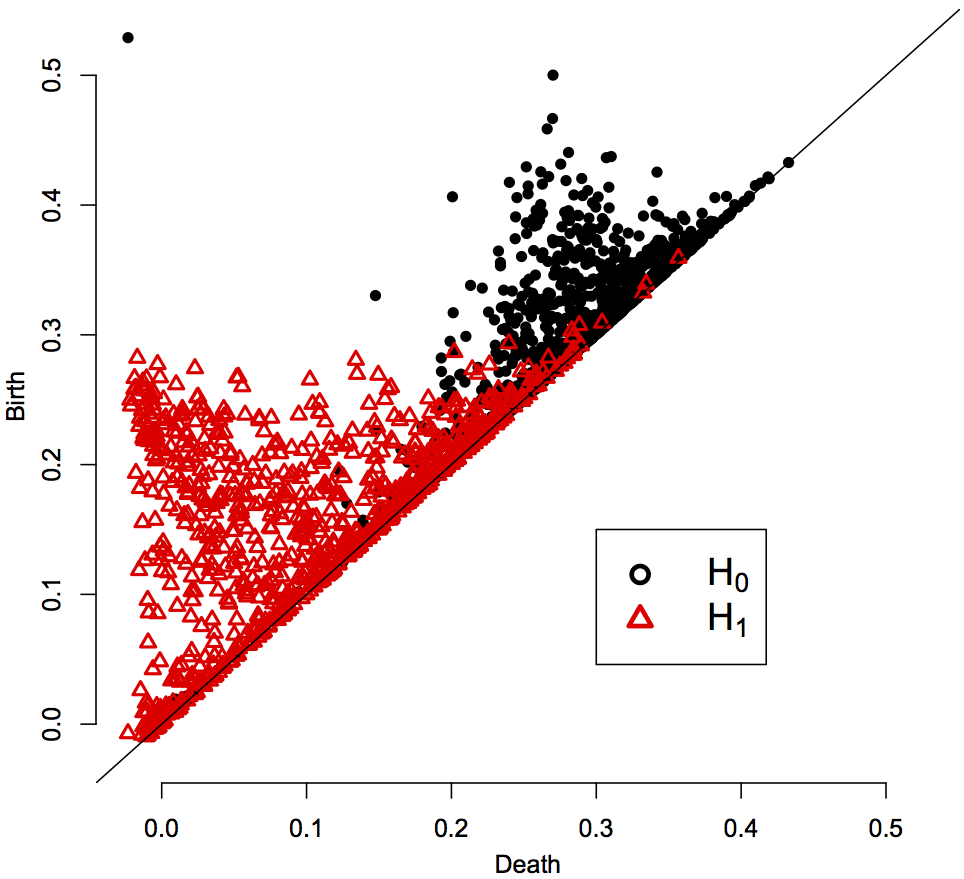}
        \caption{Human fibrin persistence diagram}\label{subfig:modeled_human_pd}
    \end{subfigure}
    \begin{subfigure}{0.49\textwidth}
        \centering
        \includegraphics[height=2in]{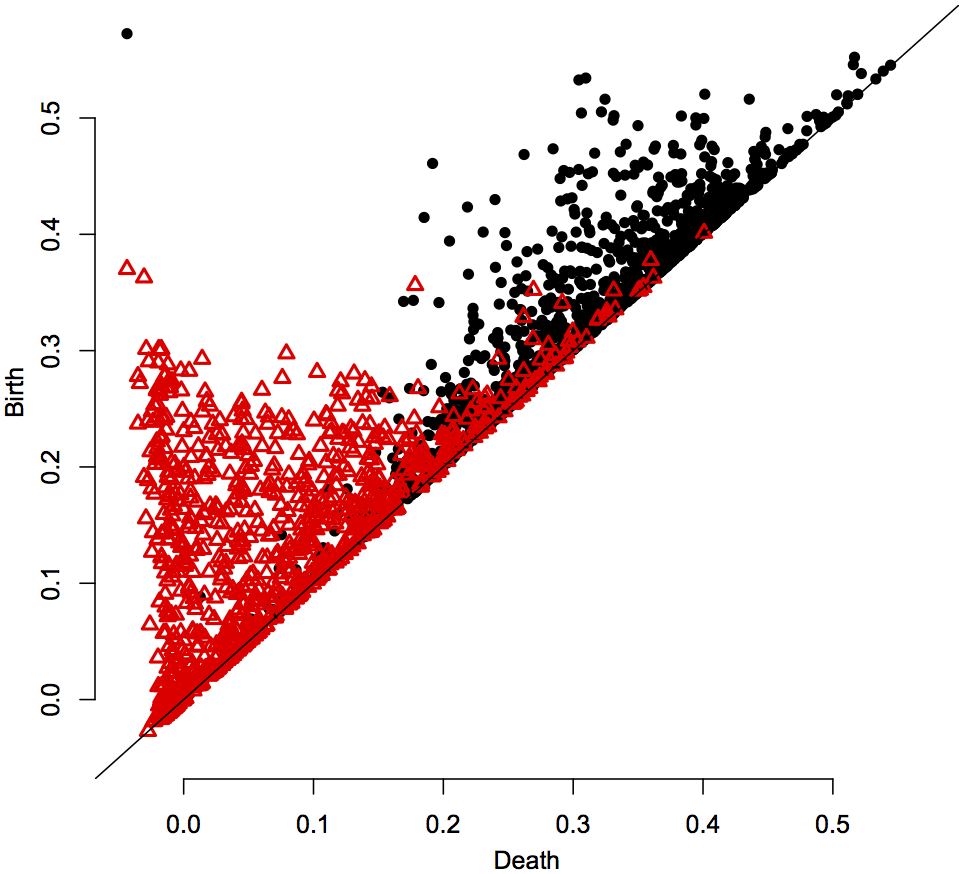}
        \caption{Monkey fibrin persistence diagram}\label{subfig:modeled_monkey_pd}
    \end{subfigure}
    \caption{Persistence diagrams for the modeled human (left) and monkey
    (right) fibrin networks displayed in \figref{human_monkey} computed using an upper-level set filtration on the modeled images. }
    \label{fig:human_monkey_pd}
\end{figure}

As noted previously, \cite{Pretorius:2009aa} carried out hypothesis tests
between fibrin networks of different species.  We use functional summaries of
persistence diagrams to carryout two-sample hypothesis tests of a human fibrin
network and a monkey fibrin network, both based on images from
\cite{Pretorius:2009aa}.  The modeled human and monkey fibrin images are
displayed in \figref{human_monkey}, and are produced by removing the white scale bar from the original images of
\cite{Pretorius:2009aa} (see \figref{fibrin_human_full} for the original human
fibrin network).  The modeling carried out is a minimal amount of smoothing to
reduce the high contrasts of the image.  Namely, local quadratic regression is
used with an adaptive bandwidth that includes the 0.1\% of the nearest neighbors
of the point of interest.  The corresponding persistence diagrams use
upper-level set filtrations on the modeled images, and are  displayed in \figref{human_monkey_pd}.  Both diagrams have similar features with $H_0$ features appearing early in the filtration, which die off and produce $H_1$ features.

Before jumping into the fibrin dataset, we carryout a simulation study using, what we refer to as, the Pickup Sticks Simulator (STIX), in order to check the performance of the proposed two-sample tests using different types of functional summaries when the ground truth is known.  The goal of STIX is to mimic some of the spatially complex features apparent in the fibrin data.

\subsubsection{Pick-up Sticks Simulation Data} \label{sec:stix_sim}
\begin{figure}
    \centering
\includegraphics[height = 2.25in]{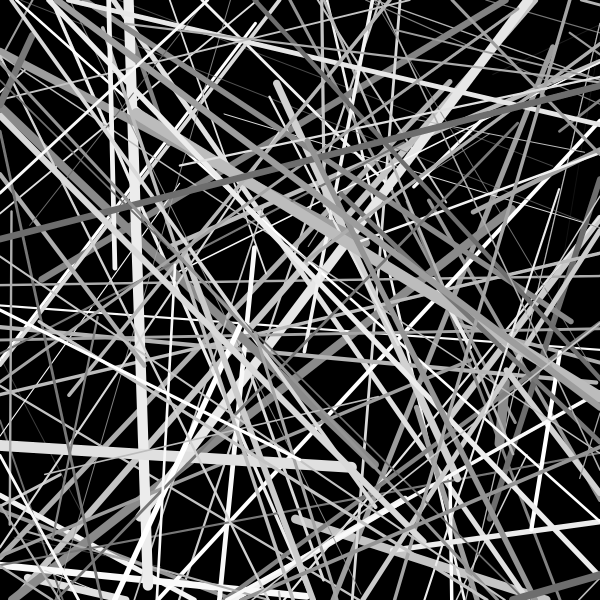} \quad \includegraphics[height = 2.25in]{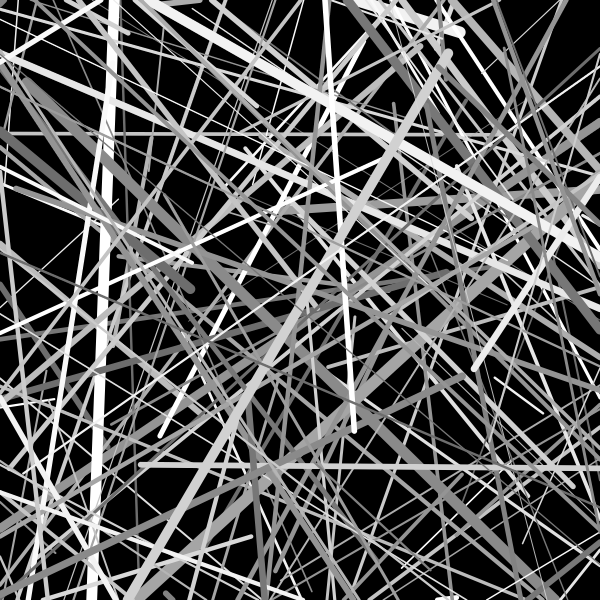}
    \caption{Realizations of the Pick-up Sticks Simulation Data (STIX) with average thicknesses of (left) 5 and (right) 6.}
    \label{fig:pickup_example}
\end{figure}

Motivated by fibrin networks, we developed data simulator that attempts to mimic some of the complicated spatial structure of fibrin.  The STIX generates data that resembles the web-like features of the fibrin networks.
The following is the STIX recipe supposing $n$ segments, or sticks, are desired
in the image.  Two sets of $n$ points are randomly sampled from a Uniform
distribution with segments drawn between points in the same position of the two
lists of random numbers.  The thickness of each segment is randomly drawn from a
$\chi^2$ distribution with thickness = $t$ degrees of freedom.\footnote{R
\citep{RCoreTeam} is used to produce the STIX images, and the thickness $t$ of
the segments are set by the {\tt{lwd}} plotting option.}  \figref{pickup_example} displays realizations of STIX with two different average thicknesses, 5 and 6.

We carryout a simulation study using STIX to check the performance of the two-sample hypothesis tests using landscape and generalized landscape functions.  Using the two-sample test framework from Section~\ref{sec:two_sample}, and Equation~\eqref{eq:two_sample} in particular, we generate images from two populations with the difference in the two populations being the average thickness of the segments via the degrees of freedom of the $\chi^2$ distribution.

For the simulation study, $\mathcal{P}_{F,1}$ is the null population, and a
thickness of~$t_1 = 5$ is used.  The alternative populations consider a range of thicknesses, $t_2 = 5, 5.25, 5.5, 5.75, 6, 7, 8$, where $t_2 = 5$ is used to check the power of the test.\footnote{This is important to ensure that the tests do not consistently \emph{incorrectly} reject the null hypothesis.}  For null thickness, $t_1$ and alternative thickness $t_2$, 100 repetitions of the following are carried out.  First, 12 STIX images are produced using both $t_1$ and $t_2$ (24 images total).  Then each image is smoothed using local quadratic regression with an adaptive bandwidth that includes the 0.1\% of the nearest neighbors, and a persistence diagram is computed using upper-level sets on the modeled images.  Then landscape functions and generalized landscapes with the triangle kernel and varying bandwidths (0.01, 0.025, 0.05, 0.10) were computed.  Permutation p-values were computed using the permutation test framework introduced in Section~\ref{sec:two_sample} using 10,000 random permutations.

The results of this simulation study for $t_2 = 5$ and $5.5$ using the 1st
homology dimension are displayed in \figref{stix_results1} with the remaining
results displayed in Appendix~\ref{app:stix}.  The median p-values or
$\log_{10}$(p-values) are displayed along with their interquartile range for 10
function orders of the landscapes and generalized landscapes considered.
\figref{stix_results1_5_500} displays the results for the case where the null and
alternative populations are the same (with an average thickness of 5), all
methods perform well with p-values distributed around 0.5.  As the alternative
population's thickness increases, all methods have p-values that get smaller.
The landscape function orders tend to have larger p-values than the generalized
landscape function orders, with the generalized landscapes with the smallest
bandwidth considered, 0.01 (red triangles), tending to have the lowest p-values
among the generalized landscapes.  The results for the case with the alternative
hypothesis average thickness of 8 are not displayed since all tests resulted in
the minimum p-value.  In Appendix~\ref{app:stix}, the simulation results for the
0th homology dimension are displayed in \figref{stix_results0} and rest
of the results for the 1st homology dimension are displayed in \figref{stix_results1_full}.

\begin{figure}
    \centering
    \begin{subfigure}{0.45\textwidth}
        \includegraphics[width = \textwidth]{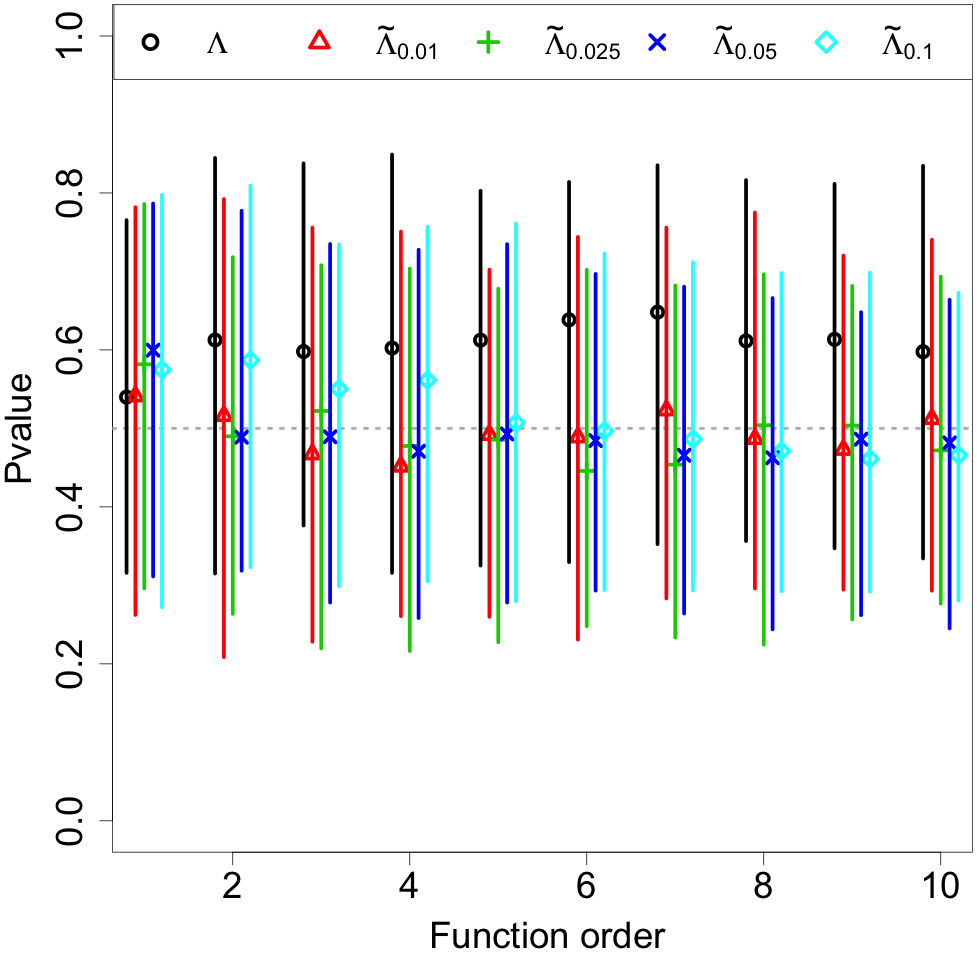}
        \caption{Homology dim=1, Thickness 5 vs.\ 5}\label{fig:stix_results1_5_500}
    \end{subfigure}
    \quad
    \begin{subfigure}{0.45\textwidth}
        \includegraphics[width = \textwidth]{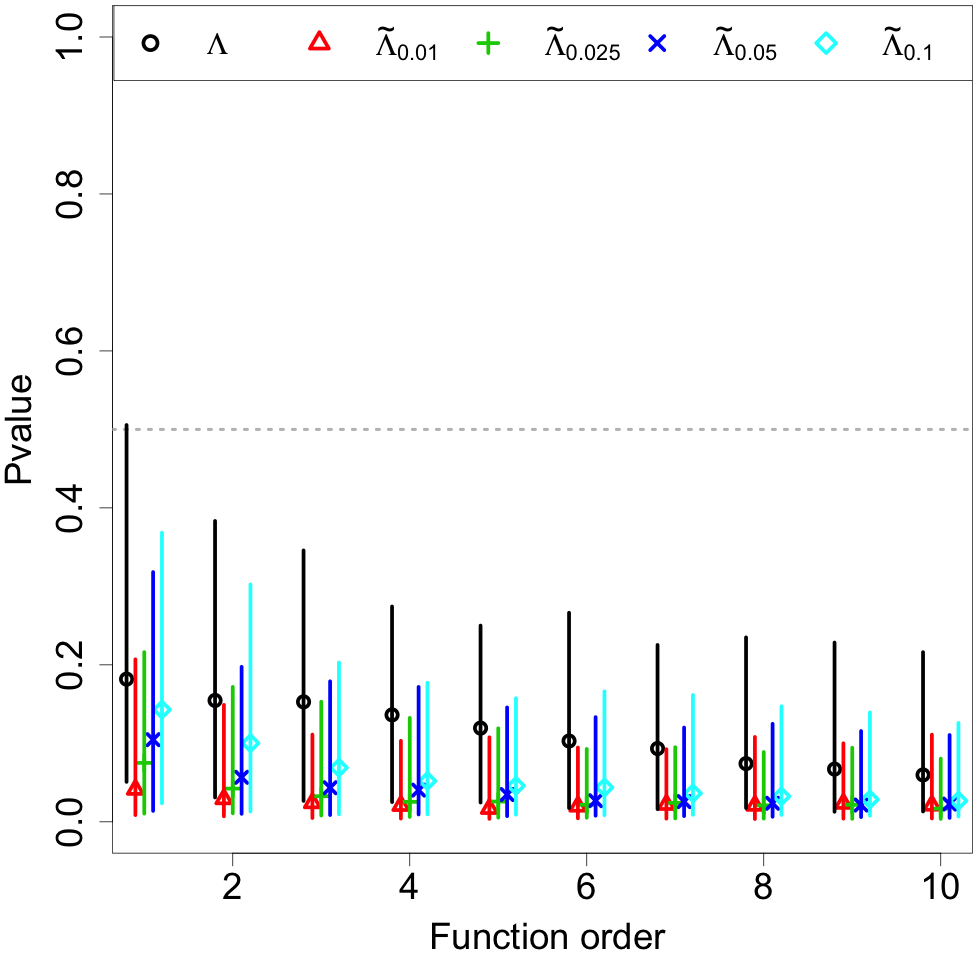}
        \caption{Homology dim=1, Thickness 5 vs.\ 5.5}\label{fig:stix_results1_5_550}
    \end{subfigure}
    \caption{STIX simulation results for Homology dimension 1.  The median permutation p-values are plotted along with their interquartile range (the vertical lines) for two-sample hypothesis tests comparing samples drawn from the null population, $\mathcal{P}_{F,1}$, with an average thickness of $t_1 = 5$.  The alternative hypotheses include average thicknesses of $t_2 = 5$ and $5.5$, corresponding to plots A and B, respectively.  The permutation p-values are based on 100 repetitions of 12 STIX images drawn from the null and alternative hypothesis, with 10,000 random permutations.
    The different plot colors and symbols represent the different functions and
    bandwidths considered; the function order is the ordering of the landscape
    and generalized landscape functions (see the discussion around
    Equation~\eqref{eq:landscape}).  The results for $t_2 = 5, 5.25, 5.5, 5.75,
    6, 7$ are displayed in \figref{stix_results1_full}.}
    \label{fig:stix_results1}
\end{figure}

\subsubsection{Fibrin Data Results}

\begin{figure}
    \centering
    \begin{subfigure}{0.45\textwidth}
\includegraphics[width = \textwidth]{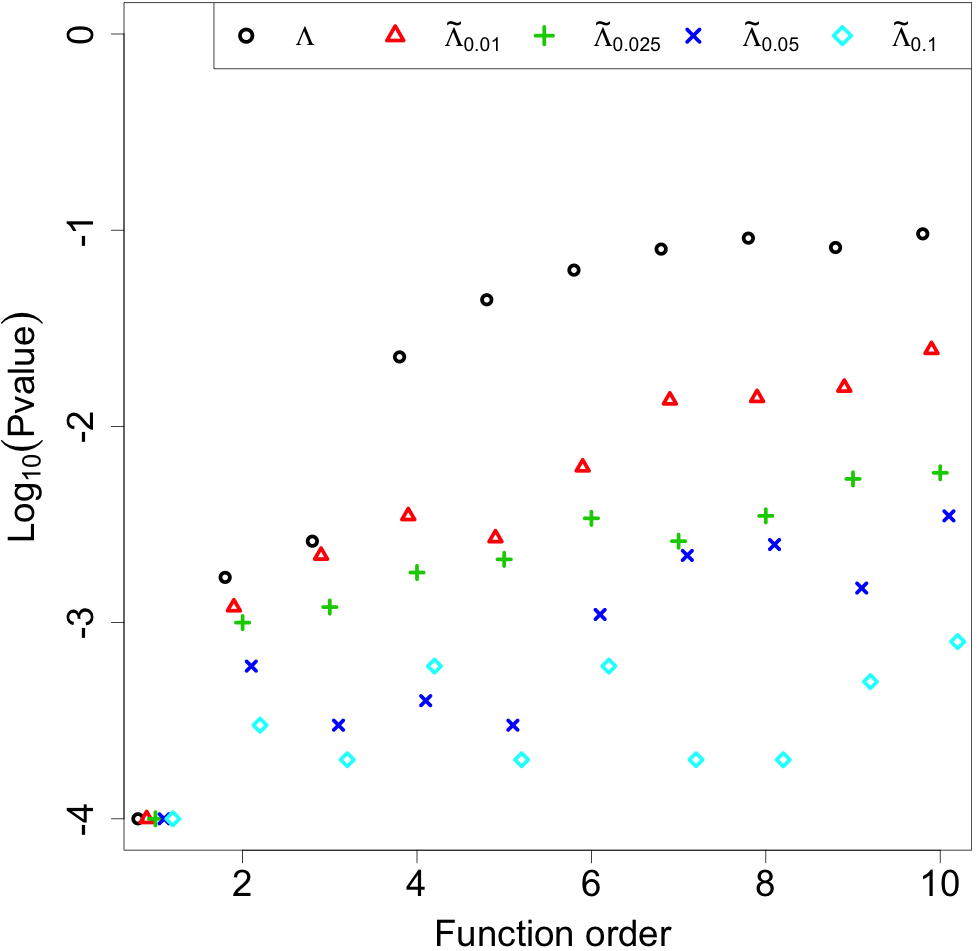}
        \caption{Homology dim=$0$}     \label{fig:results_fibrin_0}
    \end{subfigure}
    \begin{subfigure}{0.45\textwidth}
\includegraphics[width = \textwidth]{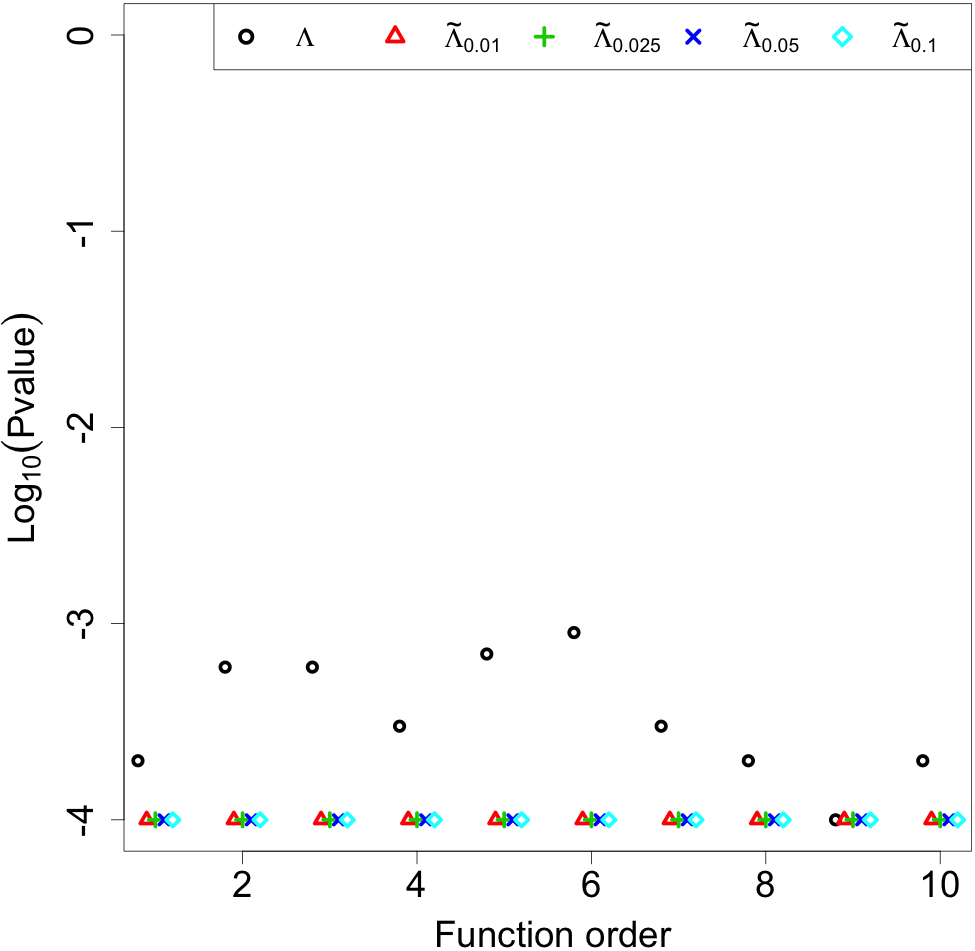}
        \caption{Homology dim=$1$}     \label{fig:results_fibrin_1}
    \end{subfigure}
\caption{Human vs.\ monkey fibrin results.  Log$_{10}$ p-values from the
    two-sample permutation hypothesis tests using the images of monkey and human
    fibrin from  \cite{Pretorius:2009aa} (a modeled version of these images are
    displayed in \figref{human_monkey}).  The human and monkey dataset were
    composed of $12$ images each (the original image divided).  The tests were carried out using 10,000 random permutations.
The different plot colors and symbols represent the different functions and
    bandwidths considered; the function order is the ordering of the landscape
    and generalized landscape functions (see the discussion around
    Equation~\eqref{eq:landscape}).
    }
    \label{fig:results_fibrin}
\end{figure}

In order to carryout a  two-sample test of the human and monkey fibrin images
from \cite{Pretorius:2009aa}, the images are first divided into 12 sub-images
(3 by 4) because only a single image of each group was available.  The
sub-images are then smoothed using local quadratic regression with an adaptive
bandwidth that includes the 0.1\% of the nearest neighbors, and a persistence
diagram is computed using upper-level sets on the modeled sub-images.  Then
landscape functions and generalized landscapes with the triangle kernel and
varying bandwidths (0.01, 0.025, 0.05, 0.10) were computed for each sub-image,
and permutation tests were carried out.  The results are displayed in \figref{results_fibrin}.  For homology dimension 0, the generalized landscapes tended to have lower p-values than the landscapes for all function orders except all methods had the minimum p-value for the first function order.  For function orders 2 - 10, the generalized landscape with the smallest bandwidth considered (0.01) tended to have the next highest p-values, and the generalized landscape with the largest bandwidth considered tended to have the lowest p-values.  For homology dimension 1, all of the generalized landscapes achieved the minimum p-value, and the landscape p-values tended to be slightly higher.

\section{Discussion}
Statistical analysis of persistence diagrams is challenging, which has led to
the development of a variety of functional transformations of the persistence
diagrams.  We reviewed the popular
functional summaries
proposed in the literature.
For the
landscape function, we generalize the formulation in order to allow more flexibility in the shape and width of the kernel, rather than requiring isosceles right triangles.

By putting analysis of persistence diagrams into a functional framework, we explained how the many tools of functional data analysis can be employed for their analysis.
We find that the average of functional summaries is a consistent estimator of the population mean function, which allows us to view the sample mean functional summary
as an estimator for this population characteristic.
We analyze some basic functional convergence theorems for the persistence functional summaries,
including a pointwise convergence and a uniform convergence theorem.
Moreover, we propose a bootstrap procedure for assessing the uncertainty of the the sample mean functional summary
and show that one can construct an asymptotically valid confidence band of the underlying population mean functional summary.
Using the proposed framework and the convergence theorems, we
show that one can conduct various statistical analyses of the data
such as constructing a prediction region for future functional summaries,
performing a two-sample test to determine if two sets of persistence diagrams are from the same population,
classifying persistence diagrams into multiple classes,
partitioning persistence diagrams into clusters,
and visualizing the relationship between several persistence diagrams.

In the simulation studies of Section~\ref{sec:sim_study}, the proposed generalized landscapes performed better than the traditional landscapes and the silhouettes in terms of test classification error for the Gleason data, and generally had lower p-values than the traditional landscapes in the two-sample hypothesis tests for the STIX simulation study (when the alternative hypothesis was true).  However, the generalized landscapes come with the cost of needing to select a kernel and,
more importantly, the bandwidth.  One benefit of the generalized landscape formulation is that more information of the persistence diagram can be packed in fewer function orders, which can aid in dimension reduction.
Before carrying out the two-sample hypothesis tests on the fibrin network images
from \cite{Pretorius:2009aa}, we developed a new data simulator, STIX, which has
a similar complicated spatial structure to the fibrin network and provided an
interesting testing ground for the proposed methods.  Without needing to measure
the widths of any of the sticks, the proposed tests based on the persistence
functional summaries were able to detect small differences in the sampled
populations \emph{when the two populations actually differed}, and detected no
difference in the case where the two samples came from the~same~population.

\paragraph*{Acknowledgments:}\
On behalf of all authors, the corresponding author states that there is no
conflict of interest.
Berry and Fasy acknowledges supported by NIH and NSF under
Grant Numbers NSF DMS 1664858 and NSF DMS 1557716.
Chen is supported by NIH grant number U01 AG016976.
Cisewski-Kehe thanks the great staff and available resources at the Yale Center for Research Computing.
Fasy is additionally supported by NSF CCF 1618605.


\bibliography{references}

\newpage

\appendix

\section{STIX simulation results} \label{app:stix}

For completeness, the full results for the STIX simulation study from
\S\ref{sec:stix_sim} are displayed in \figref{stix_results0} and
\ref{fig:stix_results1_full}.  These are the full results corresponding to the
0th and 1st homology dimension, respectively, while a subset of the results for
the 1st homology dimension are displayed in \figref{stix_results1} and discussed in the main text.

\newcommand{\figwidth}{.32}
\begin{figure}
    \centering
\begin{subfigure}{\figwidth\textwidth}
    \includegraphics[width = \textwidth]{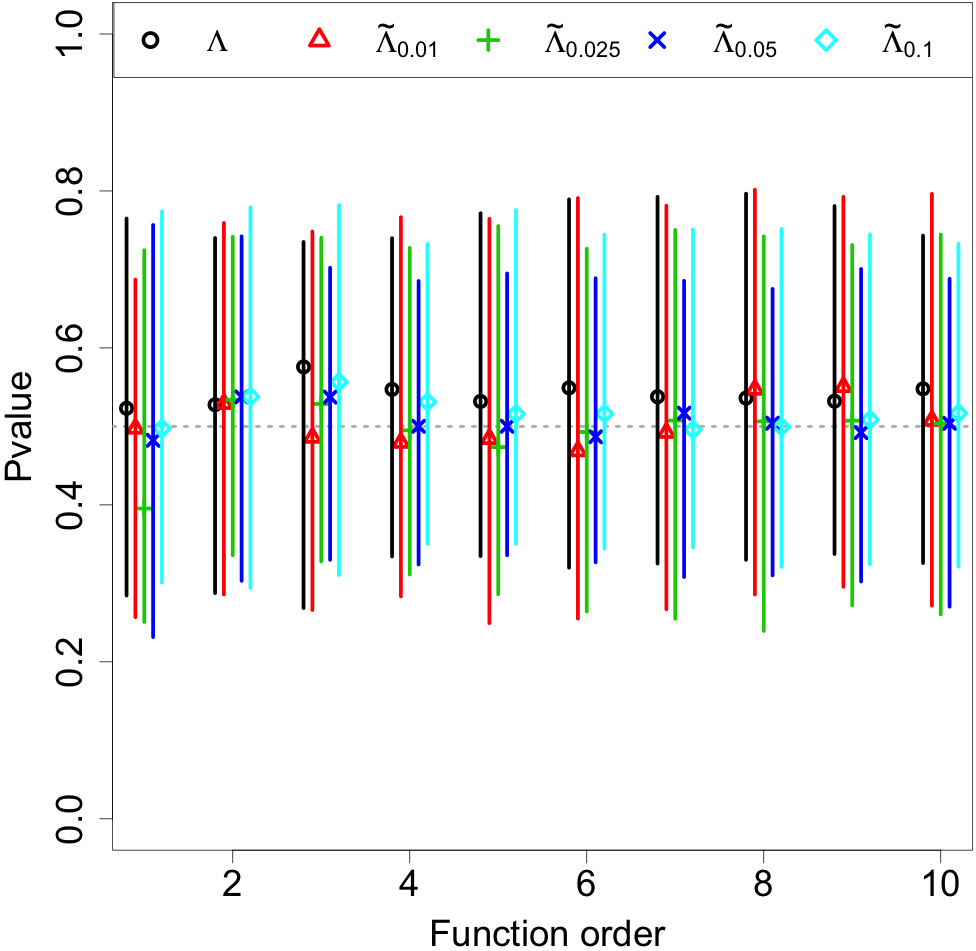}
    \caption{Hom. dim=0,  5 vs.\ 5}\label{fig:stix_results0_5_500_app}
\end{subfigure}
\begin{subfigure}{\figwidth\textwidth}
    \includegraphics[width = \textwidth]{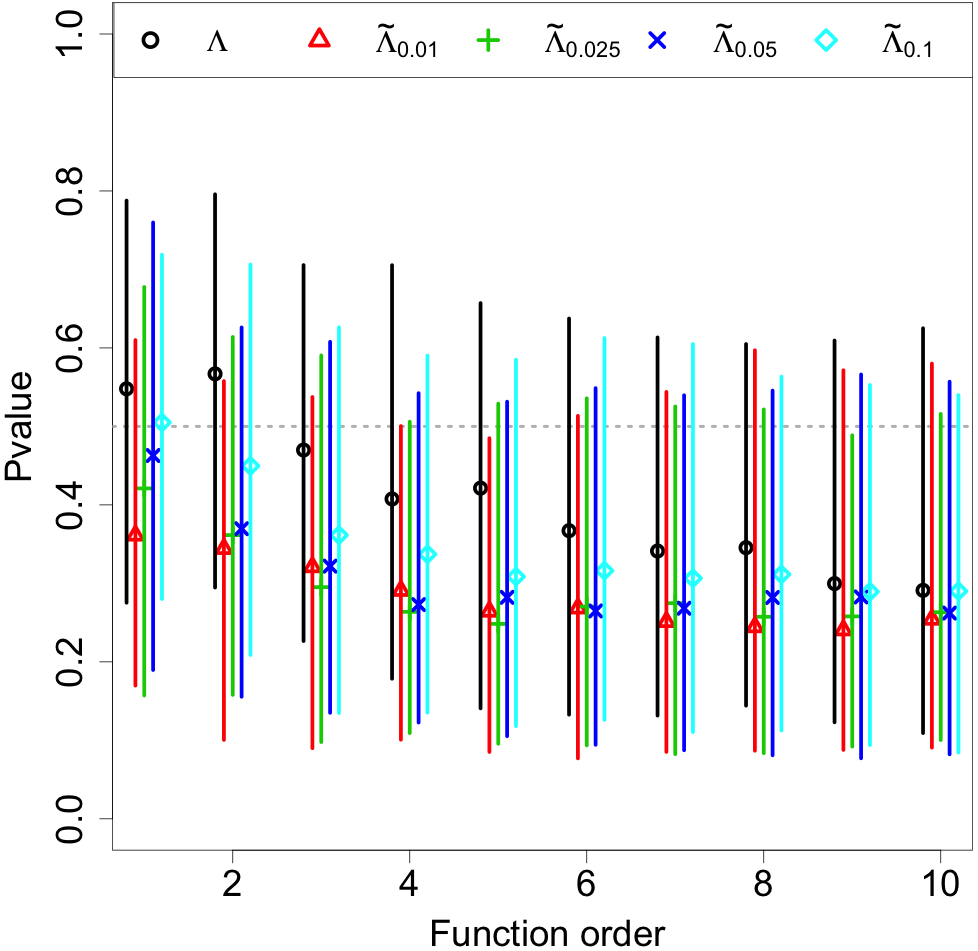}
    \caption{Hom. dim=0, 5 vs.\ 5.25}\label{fig:stix_results0_5_525_app}
\end{subfigure}
\begin{subfigure}{\figwidth\textwidth}
    \includegraphics[width = \textwidth]{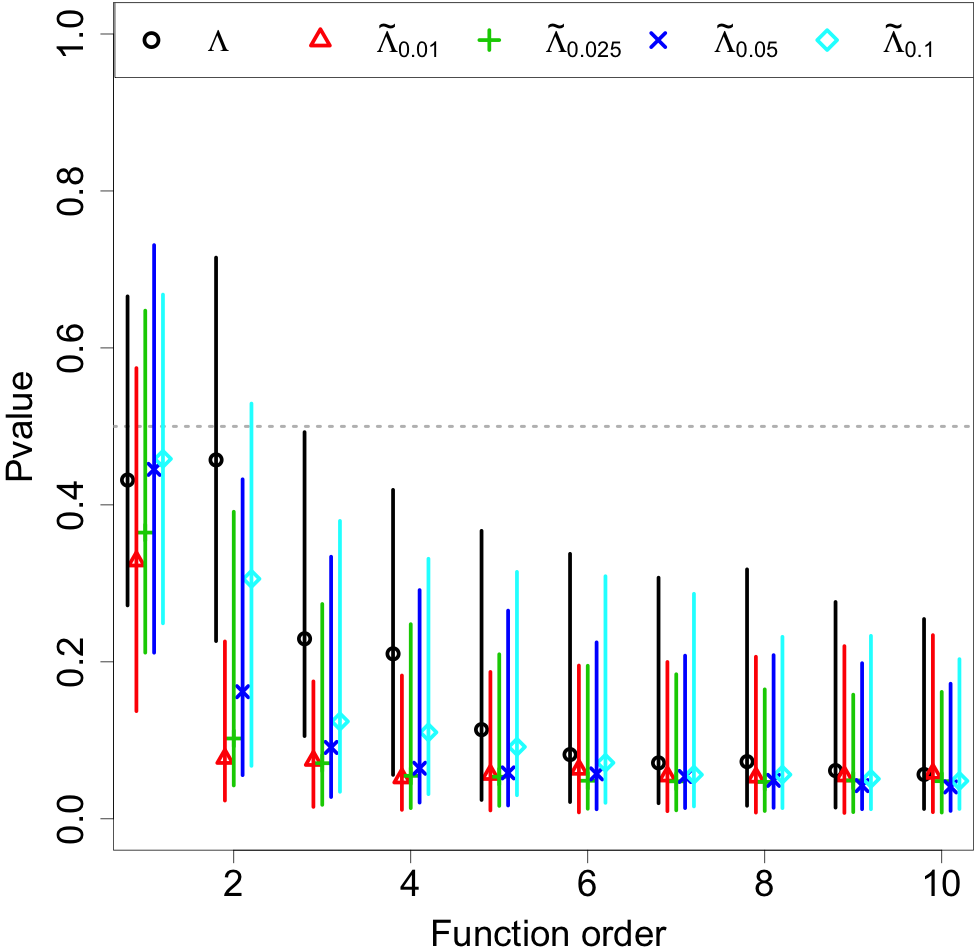}
    \caption{Hom. dim=0,  5 vs.\ 5.5}\label{fig:stix_results0_5_550_app}
\end{subfigure} \\
\begin{subfigure}{\figwidth\textwidth}
    \includegraphics[width = \textwidth]{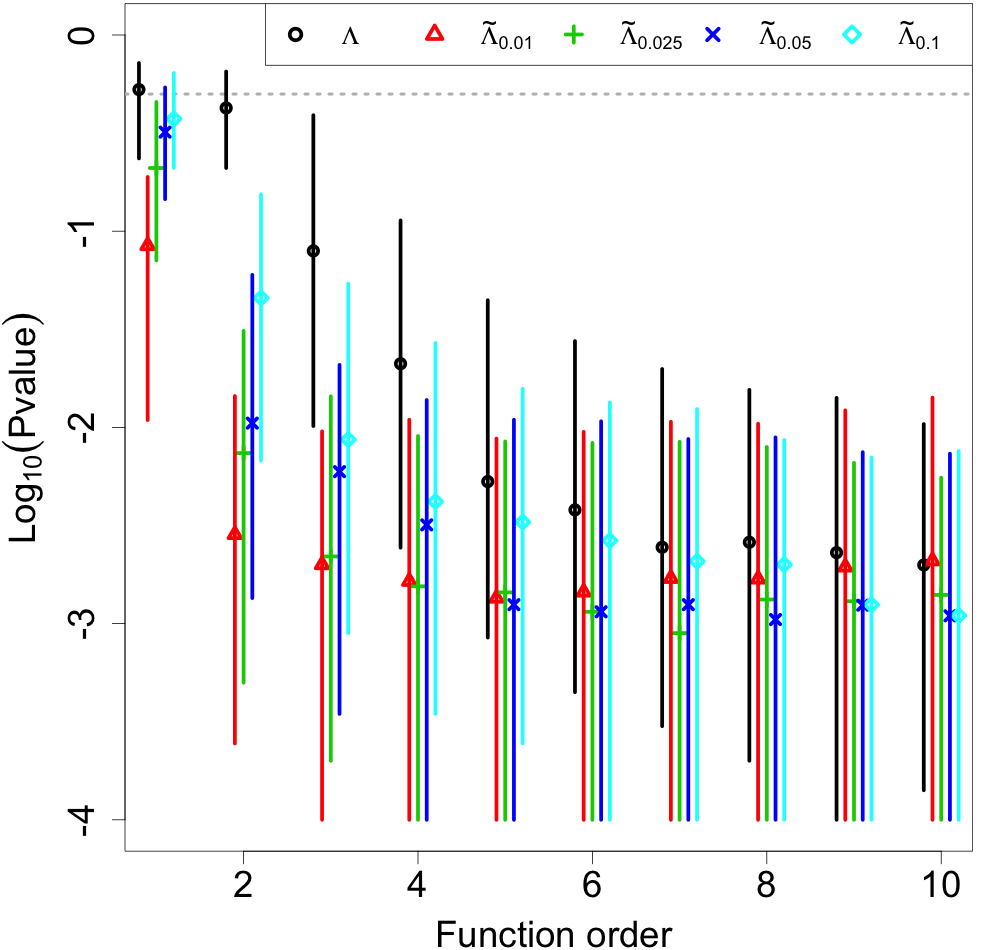}
    \caption{Hom. dim=0,  5 vs.\ 5.75}\label{fig:stix_results0_5_575_app}
\end{subfigure}
\begin{subfigure}{\figwidth\textwidth}
    \includegraphics[width = \textwidth]{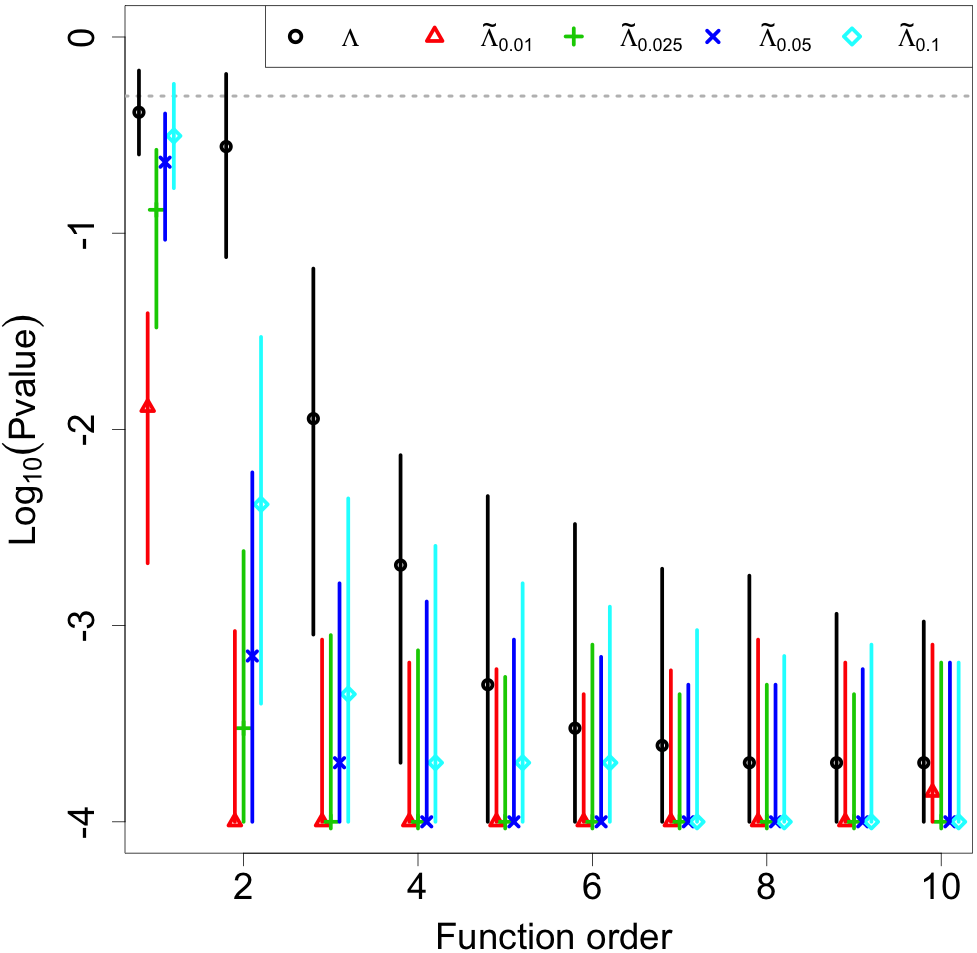}
    \caption{Hom. dim=0,  5 vs.\ 6}\label{fig:stix_results0_5_600_app}
\end{subfigure}
\begin{subfigure}{\figwidth\textwidth}
    \includegraphics[width = \textwidth]{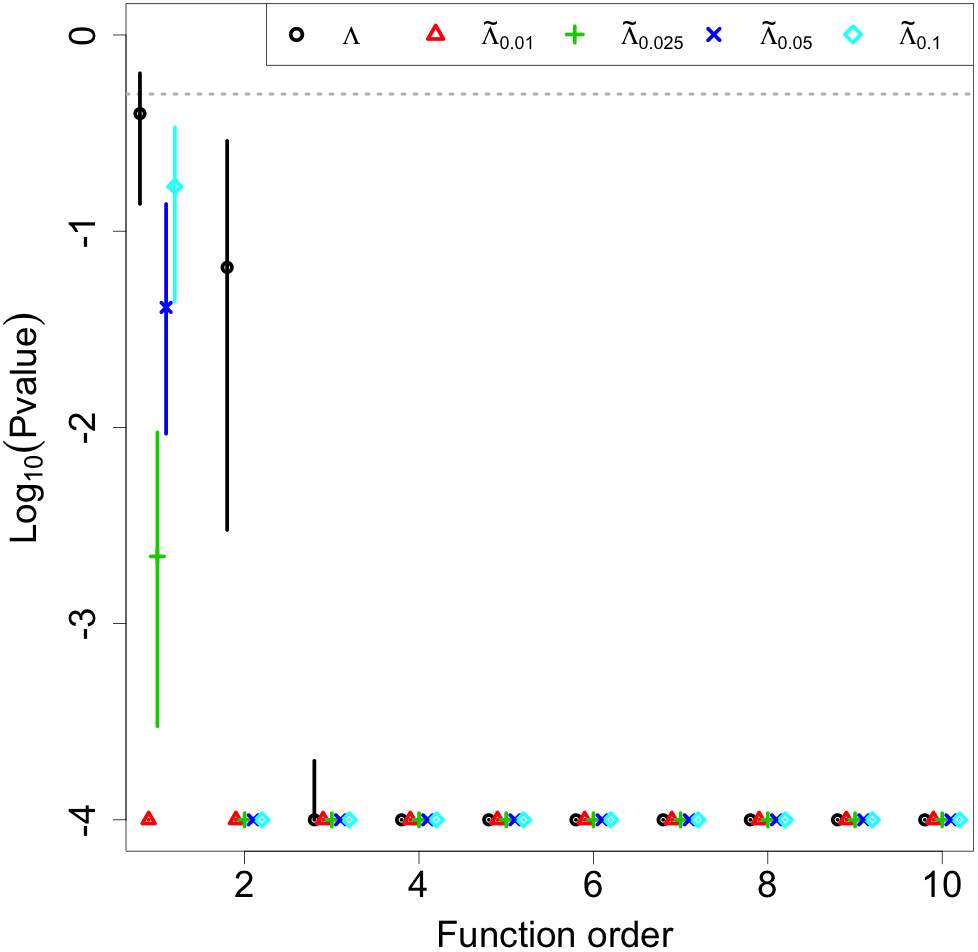}
    \caption{Hom. dim=0,  5 vs.\ 7}\label{fig:stix_results0_5_700_app}
\end{subfigure}
\caption{STIX simulation results for Homology dimension 0.  The median
    permutation p-values (a) - (c) and $\log_{10}$(pvalues) (d) - (f) are
    plotted along with their interquartile range (the vertical lines) for
    two-sample hypothesis tests comparing samples drawn from the null
    population, $\mathcal{P}_{F,1}$, with an average thickness of~$t_1 = 5$.  The alternative hypotheses include average thicknesses of $t_2 = 5, 5.25, 5.5, 5.75, 6, 7, 8$, corresponding to images (a) - (f), respectively (except average thickness of 8 is not displayed).  The permutation p-values are based on 100 repetitions of 12 STIX images drawn from the null and alternative hypothesis, with 10,000 random permutations.
The different plot colors and symbols represent the different functions and
    bandwidths considered; the function order is the ordering of the landscape
    and generalized landscape functions (see the discussion around
    Equation~\eqref{eq:landscape}).}
\label{fig:stix_results0}
\end{figure}

\begin{figure}
    \centering
\begin{subfigure}{\figwidth\textwidth}
    \includegraphics[width = \textwidth]{image1_5_500.png}
    \caption{Hom. dim=1,  5 vs.\ 5}\label{fig:stix_results1_5_500_app}
\end{subfigure}
\begin{subfigure}{\figwidth\textwidth}
    \includegraphics[width = \textwidth]{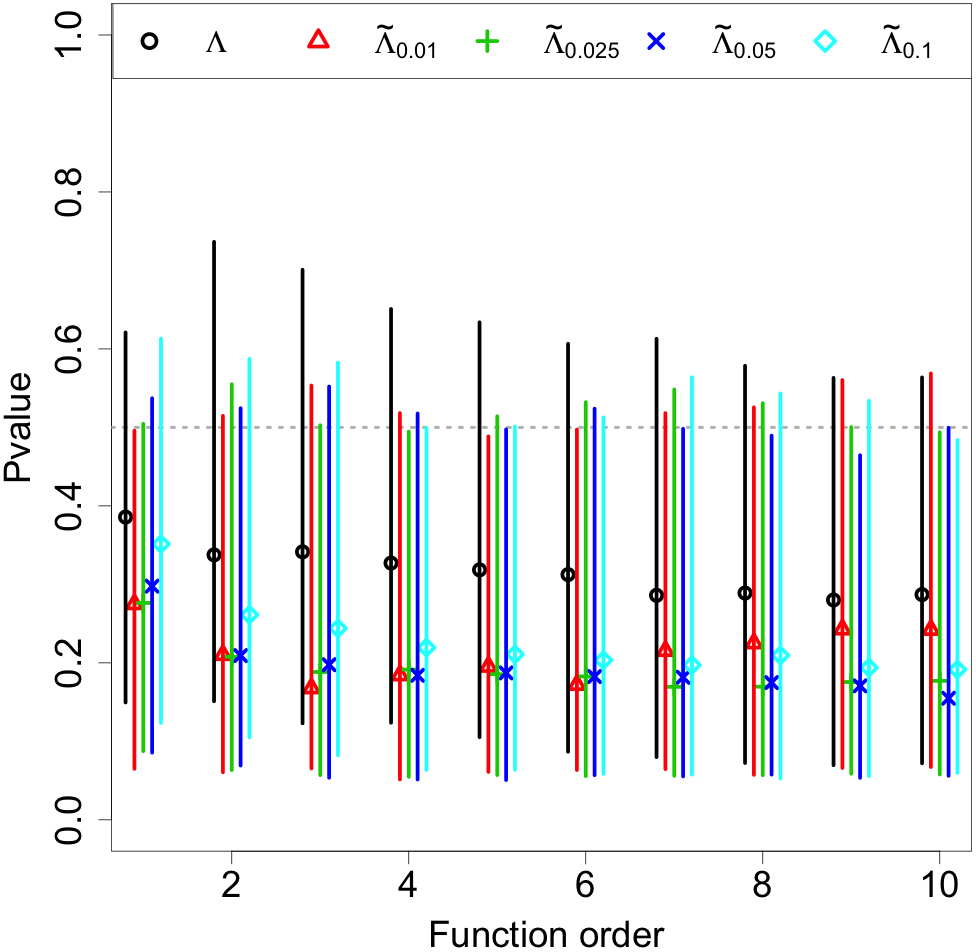}
    \caption{Hom. dim=1, 5 vs.\ 5.25}\label{fig:stix_results1_5_525_app}
\end{subfigure}
\begin{subfigure}{\figwidth\textwidth}
    \includegraphics[width = \textwidth]{image1_5_550.png}
    \caption{Hom. dim=1,  5 vs.\ 5.5}\label{fig:stix_results1_5_550_app}
\end{subfigure} \\
\begin{subfigure}{\figwidth\textwidth}
    \includegraphics[width = \textwidth]{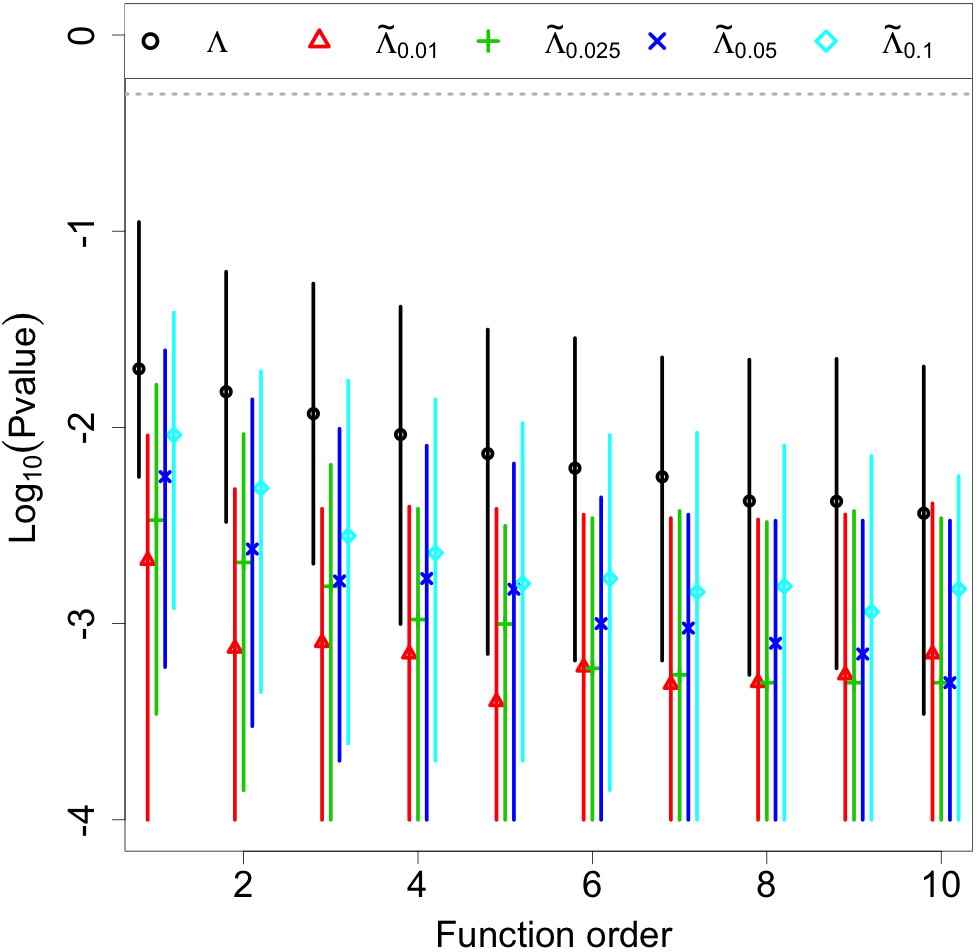}
    \caption{Hom. dim=1,  5 vs.\ 5.75}\label{fig:stix_results1_5_575_app}
\end{subfigure}
\begin{subfigure}{\figwidth\textwidth}
    \includegraphics[width = \textwidth]{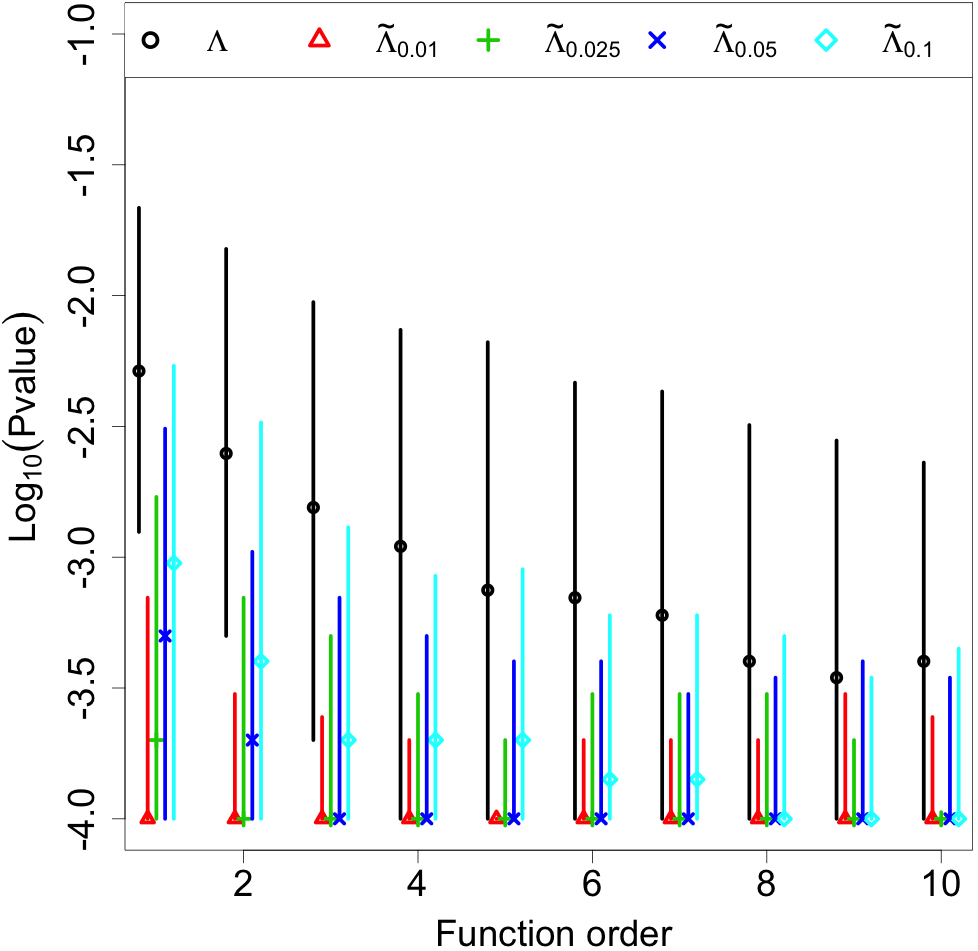}
    \caption{Hom. dim=1,  5 vs.\ 6}\label{fig:stix_results1_5_600_app}
\end{subfigure}
\begin{subfigure}{\figwidth\textwidth}
    \includegraphics[width = \textwidth]{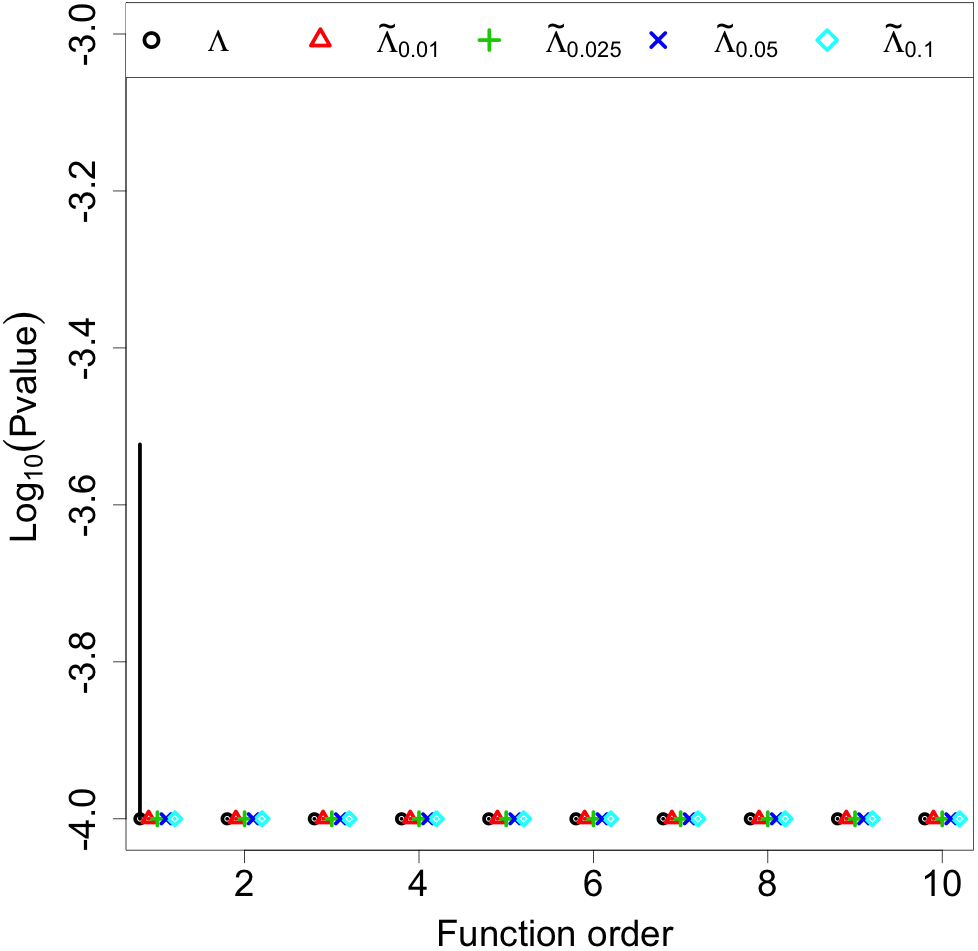}
    \caption{Hom. dim=1,  5 vs.\ 7}\label{fig:stix_results1_5_700_app}
\end{subfigure}
\caption{STIX simulation results for Homology dimension 1.  The median permutation p-values (a) - (c) and $\log_{10}$(pvalues) (d) - (f) are plotted along with their interquartile range (the vertical lines) for two-sample hypothesis tests comparing samples drawn from the null population, $\mathcal{P}_{F,1}$, with an average thickness of $t_1 = 5$.  The alternative hypotheses include average thicknesses of $t_2 = 5, 5.25, 5.5, 5.75, 6, 7, 8$, corresponding to images (a) - (f), respectively (except average thickness of 8 is not displayed).  The permutation p-values are based on 100 repetitions of 12 STIX images drawn from the null and alternative hypothesis, with 10,000 random permutations.
The different plot colors and symbols represent the different functions and bandwidths considered; the function order is the ordering of the landscape and generalized landscape functions (see the discussion around Equation~\eqref{eq:landscape}).
}
\label{fig:stix_results1_full}
\end{figure}

\section{Proofs}	\label{sec::pf}

The proofs from the propositions of Section~\ref{sec:methods} are presented below.

\begin{proof}{\bf of Proposition \ref{prop::as}.}
This proof uses ideas from
\cite{rubin1956uniform} and \cite{yuan1997theorem}.

Let $\epsilon>0$ be given.
Because $\mathcal{B}_F$ is equicontinuous,
the collection of differences
$$\mathcal{B}_{\Delta} = \{\hat{F} - \bar{F}:n = 1,2,\cdots\}$$
is also equicontinuous.
Since $\mathbb{T}$ is compact,
there exists a number $M>0$ and points $t_1,\cdots,t_M\in \mathbb{T}$
such that
$$
\sup_{\Delta\in\mathcal{B}_{\Delta}} \min_{j =1,\cdots,M}|\Delta(t)-\Delta(t_j)| < \epsilon/2.
$$
Namely, $t_1,\cdots,t_M$ forms a $\epsilon/2$-covering of $\mathcal{B}_{\Delta}$.

Let $\Delta_n(t) = \hat{F}(t)-\bar{F}(t)$, and note that $\Delta_n\in \mathcal{B}_{\Delta}$.
Then
\begin{align*}
\sup_{t\in\mathbb{T}}|\hat{F}(t)-\bar{F}(t)| & = \sup_{t\in\mathbb{T}}|\Delta_n(t)|\\
&\leq \sup_{t\in\mathbb{T}} \min_{j\in 1,\cdots,M} |\Delta_n(t)-\Delta_n(t_j)| + \sup_{j\in1,\cdots,M} |\Delta_n(t_j)|\\
&\leq \epsilon/2 + \sup_{j\in1,\cdots,M} |\Delta_n(t_j)|.
\end{align*}

By Equation~\eqref{eq::uniformbound}, every $t\in\mathbb{T}$ satisfies $\E|F_i(t)|<\infty$
so by the strong law of large number,
$$
|\Delta_n(t_j)| \overset{a.s.}{\rightarrow} 0
$$
for every $j=1,\cdots,M$
and this implies that for any $\delta>0$, there exists $N>0$ such that
$$
P\left(\sup_{n\geq N}|\Delta_n(t_j)|>\frac{\epsilon}{2M}\right)<\frac{\delta}{M}
$$
for every $j=1,\cdots,M$.

As a result,
\begin{align*}
P\left(\sup_{t\in\mathbb{T}}|\hat{F}(t)-\bar{F}(t)|>\epsilon\right)&\leq
P\left(\epsilon/2 + \sup_{j\in1,\cdots,M} |\Delta_n(t_j)|>\epsilon\right)\\
&\leq P\left(\sup_{n\geq N} \sup_{j\in1,\cdots,M}|\Delta_n(t_j)|>\frac{\epsilon}{2}\right)\\
& \leq \sum_{j=1}^M  P\left(\sup_{n\geq N}|\Delta_n(t_j)|>\frac{\epsilon}{2M}\right)\\
&\leq \delta
\end{align*}
and the result follows.

\end{proof}

\begin{proof}{\bf of Proposition \ref{prop::normal}.}

Note that by Equation~\eqref{eq::uniformbound}, the functional summary satisfies
$$
\sup_{t\in\mathbb{T}}{\sf Var}(F_i(t)) \leq \sup_{t\in\mathbb{T}}\E(F^2_i(t))\leq \bar{U}^2<\infty.
$$

{\bf Pointwise normality.}
The first assertion $\sqrt{n}(\hat{F}(t)-\bar{F}(t))\rightarrow N(0,\sigma^2(t))$
follows from the usual central limit theorem.

{\bf Normality of integrated difference.}
To show the normality for the integrated difference,
note that
\begin{align*}
\sqrt{n}\int(\hat{F}(t)-\bar{F}(t))dt & = \sqrt{n}\int\left(\frac{1}{n}\sum_{i=1}^n F_i(t) - \bar{F}(t)\right)dt\\
& = \sqrt{n} \frac{1}{n}\sum_{i=1}^n \underbrace{\int\left(F_i(t)-\bar{F}(t)\right)dt}_{=Y_i}\\
& = \sqrt{n} \bar{Y}_n,
\end{align*}
where $\bar{Y}_n = \frac{1}{n}\sum_{i=1}^n Y_i$ and
$Y_1,\cdots,Y_n$ are IID with mean $\E(Y_i) = 0$ and variance
$$
{\sf Var}(Y_i )= \int {\sf Var}(F_i(t))dt = \int \sigma^2(t)dt<\infty.
$$
Thus, by the usual central limit theorem again, we obtain the normality.

{\bf Convergence to a Gaussian process.}
The assumption in Equation~\eqref{eq::entropy}
along with Theorem 2.5 in \cite{kosorok2007introduction}
implies that the class $\mathcal{W}$
is Donsker, which implies
that the empirical process converges to a Gaussian process.

\end{proof}

\begin{proof}{\bf of Proposition \ref{prop::CB}.}
Because Equation~\eqref{eq::entropy} implies that the class $\mathcal{W}$ is Donsker,
this proposition is a well-known result in the empirical process theory.
For instance, see the discussion on page 21 of \cite{kosorok2007introduction}.
Here we briefly highlight the basic idea.

By Proposition \ref{prop::CB} and the continuous mapping theorem (see, e.g., page 16 of \cite{kosorok2007introduction}),
$$
\sqrt{n}\sup_{t\in\mathbb{T}}|\hat{F}(t)-\bar{F}(t)| \overset{D}{\rightarrow} \sup_{t\in\mathbb{T}}|\mathbb{B}(t)|.
$$
In the case of the bootstrap, using the same proposition but now apply it to the bootstrap version, we obtain
$$
\sqrt{n}\sup_{t\in\mathbb{T}}|\hat{F}^*(t)-\hat{F}(t)| \overset{D|F^{\otimes n}}{\rightarrow} \sup_{t\in\mathbb{T}}|\mathbb{B}(t)|,
$$
where $\overset{D|F^{\otimes n}}{\rightarrow}$ denotes convergences in distribution given $F_1,\cdots,F_n$.

Therefore, the bootstrap quantile converges to the corresponding quantile of the original difference.

\end{proof}

\begin{proof}{\bf of Proposition \ref{prop::pred}.}
This proof consists of two parts.
In the first part, we prove that $\hat{q}_\gamma\overset{P}{\rightarrow} q_\gamma$.
In the second part, we prove the desired result.

{\bf Part 1.}
Given $F_1,\cdots,F_n$,
we define
$$
\hat{Q}(t) = \frac{1}{n}\sum_{i=1}^n I(d(F_i,\bar{F})\leq t)
$$
to be the empirical version of $Q(t)$
and define
$$
\tilde{q}_\gamma: \hat{Q}(\tilde{q}_\gamma) = \gamma,
$$
to be the corresponding $\gamma$ quantile.

Because $\hat{Q}$ is just the empirical distribution function (EDF) of $Y_1,\cdots,Y_n$ where $Y_i = d(F_i,\bar{F})$
and $Q(t)$ is the cumulative distribution function (CDF) of $Y_i$,
$$
\sup_{t} |\hat{Q}(t)-Q(t)| = O_P\left(\frac{1}{\sqrt{n}}\right).
$$

By the mean value theorem and the fact that $\hat{Q}(\tilde{q}_\gamma) -Q(\tilde{q}_\gamma) = o_P(1)$,
\begin{align*}
\hat{Q}(\tilde{q}_\gamma) -Q(\tilde{q}_\gamma) & =Q(q_\gamma) -Q(\tilde{q}_\gamma) \\
& = Q'(q_\gamma^*) (q_\gamma-\tilde{q}_\gamma)
\end{align*}
for some $q_\gamma^* \in [q_\gamma, \tilde{q}_\gamma]$.
By assumption,
$Q'(q_\gamma^*)\geq q_0>0$
so
$$
|q_\gamma-\tilde{q}_\gamma| \leq \frac{1}{q_0} \left|\hat{Q}(\tilde{q}_\gamma) -Q(\tilde{q}_\gamma)\right| =O_P\left(\frac{1}{\sqrt{n}}\right).
$$

Now, because $\max{i=1,\cdots,n}|d(F_i,\bar{F})-d(F_i,\hat{F})| \leq d(\bar{F},\hat{F})$,
$\tilde{q}_\gamma$,
the quantile of $d(F_1,\bar{F}),\cdots, d(F_n,\bar{F})$,
and $\hat{q}_\gamma$,
the quantile of $d(F_1,\hat{F}),\cdots, d(F_n,\hat{F})$,
are bounded by
$$
|\tilde{q}_\gamma-\hat{q}_\gamma| \leq d(\bar{F},\hat{F}). 
$$
which implies
$$
|\hat{q}_\gamma - q_\gamma|=O_P\left(\frac{1}{\sqrt{n}}\right)+d(\bar{F},\hat{F}).
$$

{\bf Part 2.}
Let $$
\hat{A}_\gamma = P(F_{\sf new}\subset\hat{\mathcal{F}}_\gamma|F_1,\cdots,F_n)
=P(d(F_{\sf new}, \hat{F})\leq \hat{q}_\gamma|F_1,\cdots,F_n).
$$
be the probability of interest.
By the triangular inequality,
$$
|d(F_{\sf new}, \bar{F})-d(F_{\sf new}, \hat{F}) | \leq d(\bar{F}, \hat{F}).
$$
Thus,
\begin{align*}
A_\gamma &= P(d(F_{\sf new}, \hat{F})\leq \hat{q}_\gamma|F_1,\cdots,F_n) \\
&\geq P(d(F_{\sf new}, \bar{F})\leq \hat{q}_\gamma-d(\hat{F},\bar{F})|F_1,\cdots,F_n)\\
&= Q(\hat{q}_\gamma-d(\hat{F},\bar{F}))\\
& \geq Q\left(q_\gamma-2d(\hat{F},\bar{F})+O_P\left(\frac{1}{\sqrt{n}}\right)\right) \\
& = Q(q_\gamma) + O_P\left(\left(\frac{1}{\sqrt{n}}\right)+d(\hat{F},\bar{F})\right).\\
A_\gamma& = P(d(F_{\sf new}, \hat{F})\leq \hat{q}_\gamma|F_1,\cdots,F_n) \\
&\leq P(d(F_{\sf new}, \bar{F})\leq \hat{q}_\gamma+d(\hat{F},\bar{F})|F_1,\cdots,F_n)\\
&=Q(\hat{q}_\gamma+d(\hat{F},\bar{F}))\\
& \leq Q\left(q_\gamma+2d(\hat{F},\bar{F})+O_P\left(\frac{1}{\sqrt{n}}\right)\right) \\
& = Q(q_\gamma) + O_P\left(\left(\frac{1}{\sqrt{n}}\right)+d(\hat{F},\bar{F})\right).\\
\end{align*}
Thus,
$$
A_\gamma = P(F_{\sf new}\subset\hat{\mathcal{F}}_\gamma|F_1,\cdots,F_n)= \underbrace{Q(q_\gamma)}_{=\gamma}+ O_P\left(\left(\frac{1}{\sqrt{n}}\right)+d(\hat{F},\bar{F})\right),
$$
which completes the proof.

\end{proof}

\end{document}